\documentclass[11pt,a4paper]{article}

\usepackage{amsmath,amsfonts,graphicx,natbib,psfrag,color,bm}
\usepackage[textsize=small]{todonotes}
\usepackage{algorithm}
\usepackage{algorithmic}
\usepackage{booktabs}
\usepackage{csquotes}
\usepackage{comment}
\usepackage{url}
\usepackage{textcomp}
\usepackage{lscape}
\usepackage{longtable}
\usepackage{multirow}
\usepackage{mathtools}
\usepackage{color}

\renewcommand{\vec}[1]{\boldsymbol{#1}}
\newcommand{\indep}{\overset{indep}{\sim }}
\renewcommand{\L}{\ell}

\title{Sequential Bayesian inference for spatio-temporal models of temperature and humidity data}
\author{Yingying Lai \and\ Andrew Golightly\thanks{email: \texttt{andrew.golightly@ncl.ac.uk}} \and\ Richard Boys}
\date{School of Mathematics, Statistics and Physics, Newcastle University,\\
  Newcastle upon Tyne, NE1 7RU, UK}

\begin{document}
\maketitle
\begin{abstract}
  We develop a spatio-temporal model to forecast sensor output at five
  locations in North East England. The signal is described using
  coupled dynamic linear models, with spatial effects specified by a
  Gaussian process. Data streams are analysed using a stochastic
  algorithm which sequentially approximates the parameter posterior
  through a series of reweighting and resampling steps. An iterated
  batch importance sampling scheme is used to circumvent particle
  degeneracy through a resample-move step. The algorithm is modified
  to make it more efficient and parallisable. The model is shown to
  give a good description of the underlying process and provide
  reasonable forecast accuracy.
\end{abstract}


\section{Introduction}
\label{sec:intro}

Climate is one of the most important environmental factors which plays
a critical role on the global mission of urban sustainability.
Consequently, it has attracted tremendous attention from academic
scientists and industrial experts in recent decades. In this paper we
focus on understanding the relationship between temperature and
humidity, as these are two of the most important factors in driving
other climate processes. Our primary objective is the development of
dynamic models which can be used to understand the stochastic nature
of temperature and humidity, as well as quantify their spatial
dependencies.  Moreover, in order to facilitate accurate forecasts in
real time, we focus on developing algorithms which allow inferences to
made sequentially.

The literature contains several temporal models for temperature at a
single location. For example, \cite{Campbell05} proposed an
autoregressive (AR) model with Fourier components to account for
seasonality, a polynomial deterministic trend and a generalised
autoregressive conditional heteroscedasticity (GARCH) error process.
Further AR modelling approaches have been proposed by \cite{Hardle11},
\cite{Benth07} and \cite{Benth12}, with the latter adopting a
continuous-time approach. Although generic approaches for spatial data
sets are widely available (see e.g. \cite{Cressie93}, \cite{Stein99},
\cite{Ripley04}, \cite{Diggle06}, \cite{Gelfand10}, \cite{Cressie11}
and \cite{Banerjee14}), relatively few papers have addressed the joint
modelling of temperature and humidity at multiple locations.
\cite{Hu13, Hu15} use a stochastic partial differential equation
(SPDE) to model yearly temperature and humidity data at 120 locations
and perform fully Bayesian inference via an integrated nested Laplace
approximation \citep{Rue09}.

The modelling approach developed here is motivated by the fine scale
temporal nature of the available data. Dynamic linear models (DLMs)
are widely used for system evolution learning and short term
forecasting due to their simple and practical structures; see, for
example, \cite{West99} for an introduction. We exploit these
properties here by specifying a marginal DLM for temperature and a
conditional DLM for humidity given temperature. We account for spatial
dependence at nearby locations by adding a spatial Gaussian process to
the system equations, thereby smoothing spatial deviations from the
underlying temporal model. A similar approach was used by
\cite{Shaddick02} for pollutant data.

We perform fully Bayesian inference for the model parameters as each
observation becomes available. Since the posterior distribution is
intractable, we use sequential Monte Carlo (SMC) methods that
approximate the posterior distribution at each time point through a
set of weighted samples; see \cite{fearnhead18} for a recent review of
SMC methods.  Although the posterior is intractable, the observed data
likelihood is available in closed form, allowing the implementation of
the iterated batch importance sampling (IBIS) scheme, first introduced
by \cite{Chopin02}; see also \cite{Chopin13} for a related approach.
Essentially, parameter samples (known in this context as particles)
are incrementally weighted by the observed data likelihood
contribution of the currently available observation. Particle
degeneracy is mitigated via a resample-move step \citep{gilks2001}
which `moves' each parameter particle through a Metropolis-Hastings
kernel that leaves the target invariant. This step can be executed
subject to the fulfilment of some degeneracy criterion e.g. small
effective sample size. However, the computational cost of the
resample-move step increases as the algorithm includes more data, as
it requires calculation of the observed data likelihood of all
available information. To obtain an online IBIS algorithm, where the
computational cost of assimilating a single observation is bounded, we
modify the resample-move step by basing the observed data likelihood
on an observation window whose length is a tuning parameter, chosen to
balance accuracy and computational efficiency. We use a simulation
study to formulate practical advice on how to choose the size of this
window.

Further computational savings can be made by employing a high
performance computing system. Whilst the weighting and move steps can
be performed independently for each particle, a basic implementation
of the resampling step requires collective operations, such as adding
up the particle weights. Our approach is to use a simple strategy
which performs the resampling step independently for batches of
parameter samples, thus allowing a fully parallel (per parameter
batch) implementation of the algorithm to be performed. We quantify
the effect of the approximation induced by this approach using
synthetic data. Finally, we apply the online IBIS scheme (with
parallel implementation) to the observed dataset and examine the model
reliability and forecast accuracy through comparison of observed
measurements with their posterior predictive distribution.

The remainder of the paper is organised as follows. A brief
description of the data is given in Section~\ref{sec:data}. The
structures of the spatial DLMs for temperature and humidity are
discussed in Section~\ref{sec:dlm}. In Section~\ref{sec:bayes}, we
introduce the IBIS scheme and develop a faster online version and then
compare the performance of both schemes in Section~4 via a simulation
study. In Section~\ref{app}, we report the full analysis on our North
East dataset on temperature and humidity and draw conclusions in
Section~\ref{disc}.

\subsection{Data collection}
\label{sec:data}

\begin{figure}[t]
\centering
\includegraphics[width=0.8\textwidth,height=0.4\textheight]{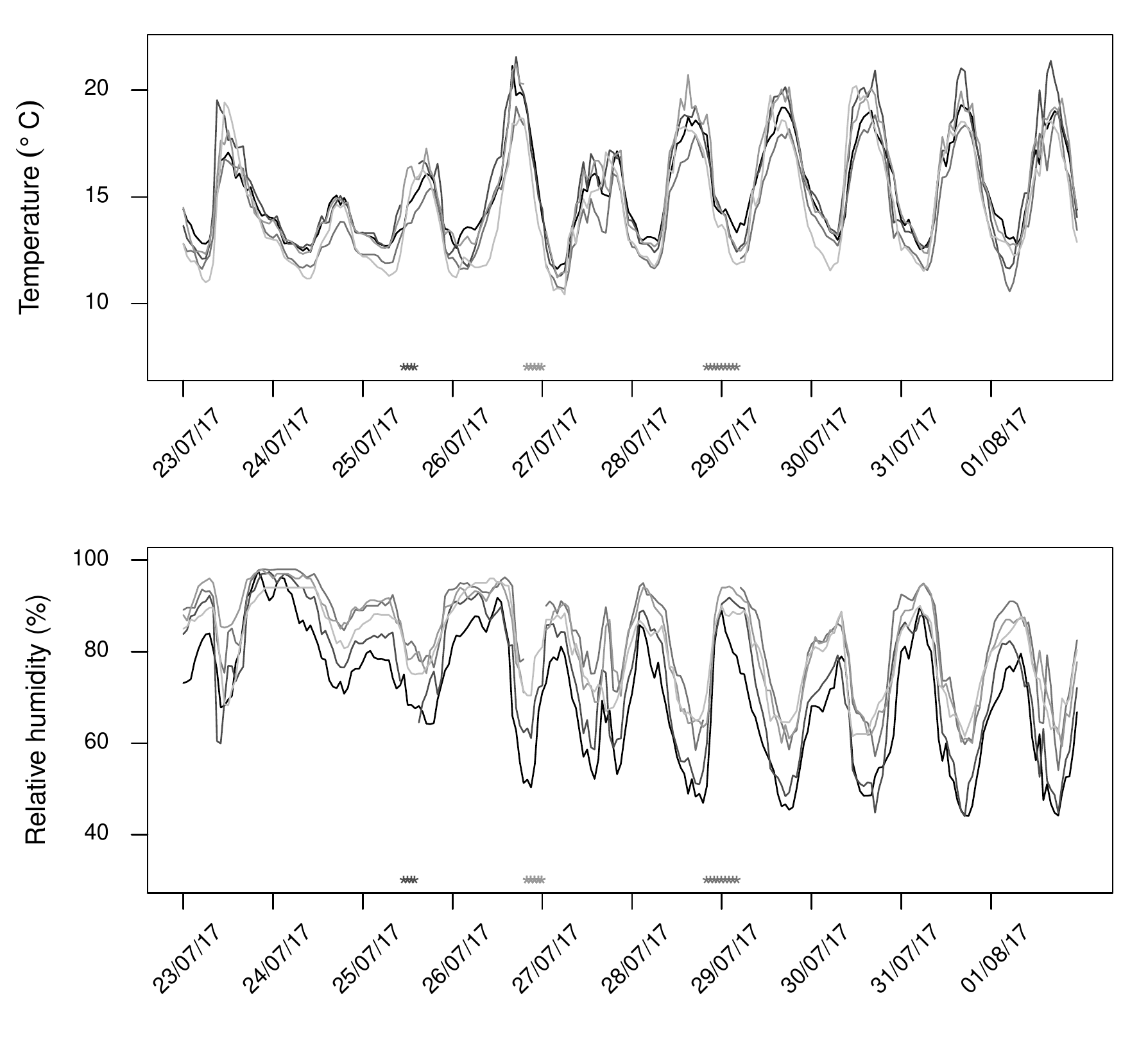}
\caption{Temperature and relative humidity data streams over time at
  each location. Periods of missingness are indicated just above the
  x-axis.}
\label{realdata}
\end{figure}

\begin{figure}[t]
\centering
\includegraphics[scale=0.8]{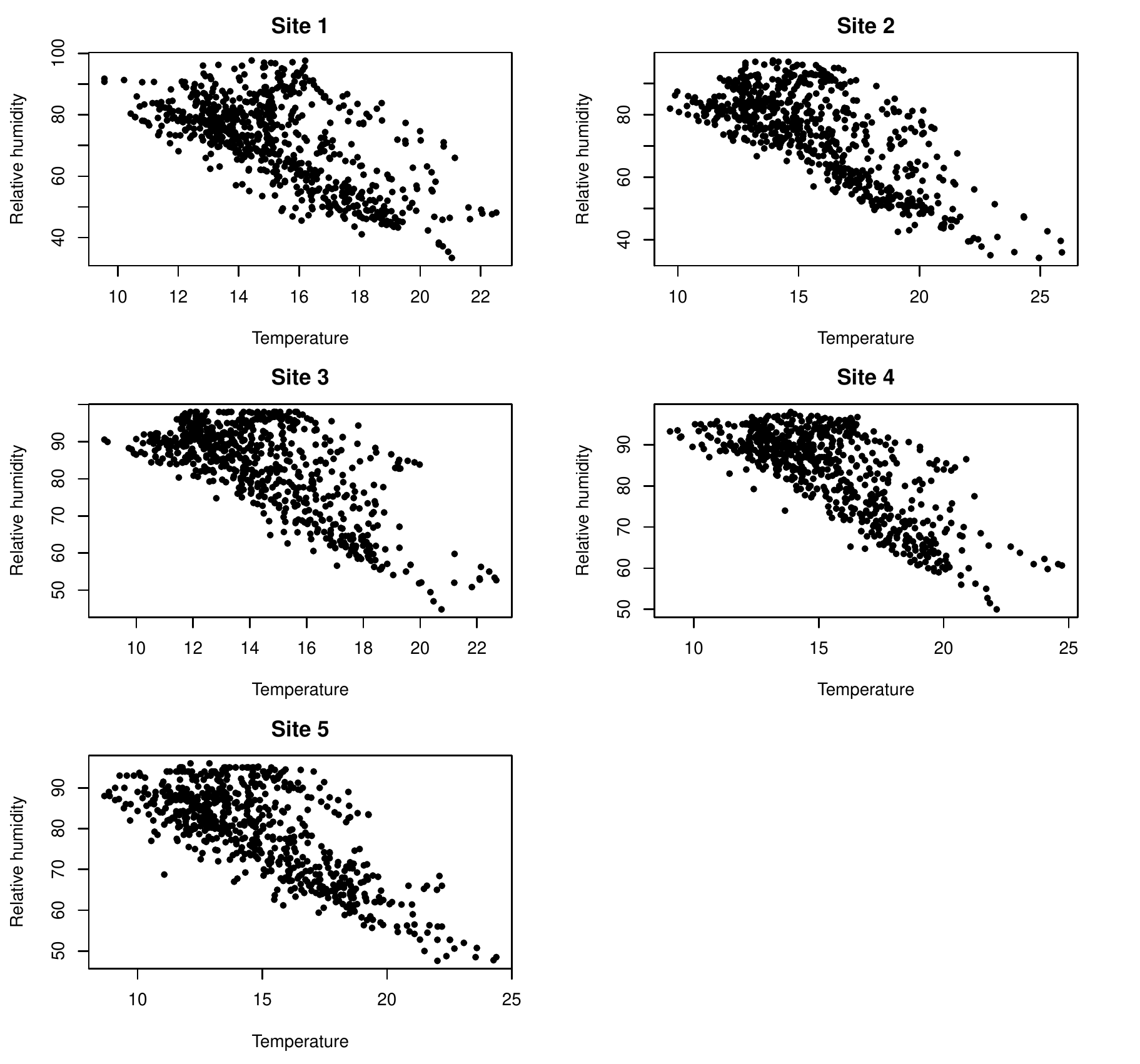}
\caption{Scatter plots of temperature against relative humidity at each location.}
\label{170610-scatter}
\end{figure}

Recent advances in sensor technology and data management mean that it
is now possible to reliably and affordably collect data on many
aspects of city life. The temperature and relative humidity data
analysed in this paper were collected from the Urban Observatory
\citep{James14}, a big data hub providing smart-city data via a grid
of sensors in North East England.  The data are received in real time,
and this requires efficient network transmission and data storage
solutions. Temperature is measured in degree Celsius, and relative
humidity is measured as the ratio of the amount of water vapour held
in the air against the the maximum amount of water vapour the air can
hold at a specific temperature. The data are captured and processed
through a microprocessor inside a sensor and transmitted via a high
speed network to the database \citep{Galatioto14}. We consider data
streams at five locations: Newcastle upon Tyne, Seaham, Peterlee,
Whitley Bay and Consett. The observation period is from 8th July 2017
to 31st December 2017. Due to the different recording frequencies of
some of the sensors, we take the average values of temperature and
relative humidity over every consecutive hour, giving a total of 4239
time points at which at least one location has a measurement.
Figure~\ref{realdata} shows the multiple data streams over time at
different locations. Both temperature and relative humidity exhibit a
clear sinusoidal pattern over each 24~hour period. Scatter plots of
humidity against temperature for each location are shown in
Figure~\ref{170610-scatter} and reveal a strong negative linear
correlation. Unfortunately, missing data are inevitable due to network
disconnection or sensor failure. Table~\ref{missingdata} and
Figure~\ref{realdata} summarise and display the proportion of missing
data at each location during the observation period.

\renewcommand{\arraystretch}{1.2}
\begin{table}[]
\centering
\label{missingdata}
\begin{tabular}{ccrrrrrcrr}
\toprule 
Variable    & Location  & \multicolumn{1}{c}{Missing} & \multicolumn{1}{c}{Prop.} & \multicolumn{1}{c}{Mean} & \multicolumn{1}{c}{Min.} & \multicolumn{1}{c}{25\%} & \multicolumn{1}{c}{Median} & \multicolumn{1}{c}{75\%} & \multicolumn{1}{c}{Max.} \\  \toprule
Temperature & Newcastle & 392                         & 9.25\%                        & 10.62                    & -9.10                    & 6.70                     & 11.70                      & 14.88                    & 27.53                    \\
(\textdegree{}C)    & Seaham    & 54                          & 1.27\%                        & 11.48                    & -2.17                    & 8.12                     & 12.30                      & 15.07                    & 25.90                    \\
            & Peterlee  & 46                          & 1.09\%                        & 10.49                    & -2.24                    & 7.37                     & 11.52                      & 13.95                    & 22.68                    \\
            & Whitley Bay   & 6                           & 0.14\%                        & 11.07                    & -4.62                    & 7.72                     & 12.10                      & 14.73                    & 24.73                    \\
            & Consett   & 306                         & 7.22\%                        & 10.40                    & -3.37                    & 6.90                     & 11.20                      & 14.24                    & 24.38                    \\ [0.3cm]
Humidity  & Newcastle & 392                         & 9.25\%                        & 83.33                    & 42.50                    & 78.33                    & 85.50                      & 90.67                    & 99.00                    \\
 (\%)          & Seaham    & 54                          & 1.27\%                        & 73.62                    & 34.23                    & 67.08                    & 74.50                      & 81.67                    & 97.42                    \\
            & Peterlee  & 46                          & 1.09\%                        & 84.86                    & 44.83                    & 80.22                    & 86.83                      & 91.67                    & 99.00                    \\
            & Whitley Bay   & 6                           & 0.14\%                        & 86.25                    & 50.00                    & 82.25                    & 88.25                      & 93.00                    & 98.25                    \\
            & Consett   & 306                         & 7.22\%                        & 83.59                    & 46.40                    & 79.33                    & 86.00                      & 90.50                    & 97.00                    \\ \toprule
\end{tabular}
\caption{A summary of hourly average temperature and humidity data
  over the period 8th July 2017 to 31st December 2017 at five
  locations in North East England.} 
\end{table}

\section{Spatial dynamic linear model (DLM)}
\label{sec:dlm}

We develop a joint model for hourly average temperature and humidity,
recorded at each of $\L$ locations. The model is specified through a
marginal model for temperature and a conditional model for humidity
given temperature. Let $\vec{X}_{t_i}=(X_{t_i}^1,\ldots,X_{t_i}^{\L})^T$
denote hourly average temperature taken over intervals
$(t_i,t_{i+1}]$, with $t_i$ in hours ($i=1,\ldots,n$) and
$\vec{Y}_{t_i}=(Y_{t_i}^1,\ldots,Y_{t_i}^{\L})^T$ denote the
corresponding humidity values. In what follows we scale time so that
$t_1=0$.

\subsection{Spatial temperature DLM}
\label{sec:tdlm}

In Section~\ref{sec:data} we noted that the data show clear
seasonality in both temperature and humidity measurements. This
suggests that marginally each variable should be modelled by a
sinusoidal form with a 24 hour period.  For simplicity, consider first
a single location $j$. We propose a DLM for temperature with
observation equation
\begin{equation}\label{obsx}
X_{t_i}^j= \vec{F}_{t_i}^{x,j}\vec{\theta}_{t_i}^{x,j}+v_{i}^{x,j},\qquad v_{i}^{x,j}\indep N(0,V^{x,j}),
\end{equation} 
where the observation matrix $\vec{F}_{t_i}^{x,j}=(\cos(\pi t_i/12),
\sin(\pi t_i/12), 1)$ and
$\vec{\theta}_{t_i}^{x,j}=(\theta_{t_i,1}^{x,j},\theta_{t_i,2}^{x,j},\theta_{t_i,3}^{x,j})^T$.
Note that, after dropping the superscripts for simplicity, the
observation equation can be written as
\begin{equation}
\label{obsx2}
X_{t_i}= \tilde{\theta}_{t_i,2}\cos\left(\frac{\pi t_i}{12}-\tilde{\theta}_{t_i,1}\right)+\theta_{t_i,3}+v_{i}
\end{equation}
where the parameters in (\ref{obsx}) and (\ref{obsx2}) are related
using
\begin{equation}\label{relation}
\tilde{\theta}_{t_i,1}=\sqrt{\theta_{t_i,1}^{2}+\theta_{t_i,2}^{2}},\qquad 
\tilde{\theta}_{t_i,2}=\tan^{-1}\left(\frac{\theta_{t_i,2}}{\theta_{t_i,1}}\right).
\end{equation}
We allow amplitude, phase shift and basal temperature to be time-varying, and take a system equation of the form
\begin{equation}\label{sysx}
\vec{\theta}_{t_i}^{x,j}=\vec{G}^{x,j}_{t_i}\vec{\theta}_{t_{i-1}}^{x,j}+k_{i}\vec{w}_{i}^{x,j}+\vec{p}_i^{x,j},\qquad \vec{w}_{i}^{x,j}\indep N\left\{\vec{0},\textrm{diag}(\vec{W}^{x,j})\right\}
\end{equation}
where the system matrix $\vec{G}^{x,j}_{t_i}=\mathbb{I}_3$, the
$3\times 3$ identity matrix, and
$\vec{W}^{x,j}=(W_{1}^{x,j},W_{2}^{x,j},W_{3}^{x,j})^T$. Note that
including~$k_i$, where $k_i^2=t_i-t_{i-1}$, allows for measurements to
be on an irregularly spaced temporal grid. Further the terms
$\vec{p}_i^{x,j}=(p_{i,1}^{x,j},p_{i,2}^{x,j},p_{i,3}^{x,j})^T$ allow
for spatial variability between amplitude, phase shift and basal
temperature values at nearby locations.  We model the components of
the spatially smooth error process $\vec{p}_i^{x,j}$ using independent
zero mean Gaussian process (GP) priors with covariance functions
$f_m^x(\cdot), m=1,2,3$, that is,
\[
p_{i,m}^{x,j}\sim GP\{\vec{0},f_m^x(\cdot)\},\quad m=1,2,3.
\] 
We take these covariance functions to have a simple exponential form
\[
f_m^x(d_{jj'})=\textrm{Cov}(\theta_{t_i,m}^{x,j},\theta_{t_i,m}^{x,j'})
=\sigma_{x,m}^2\exp(-\psi_{x,m} d_{jj'}),\quad m=1,2,3
\]
and depend on parameters $\vec{\sigma}_x = (\sigma_{x,1},
\sigma_{x,2}, \sigma_{x,3})$ and $\vec{\psi}_x = (\psi_{x,1},
\psi_{x,2}, \psi_{x,3})$, with the latter determining the decay ratio
of the correlation as the distance between two locations $d_{jj'}$
increases \citep{Banerjee14}.

The full spatial DLM model (over all locations) can be written as
\begin{equation}
\begin{split}
\vec{X}_{t_i} &= \vec{F}_{t_i}^{x}\vec{\theta}_{t_i}^{x}+\vec{v}_{i}^{x},\qquad \vec{v}_{i}^{x}\indep N\{\vec{0},\textrm{diag}(V^{x,1},\ldots,V^{x,\L})\}, \\  
\vec{\theta}_{t_i}^x &=\vec{\theta}_{t_{i-1}}^{x}+k_{i}\vec{w}_{i}^{x}+\vec{p}_i^x,\qquad \vec{w}_{i}^{x}\indep N\{\vec{0},\textrm{diag}(\vec{W}^{x,1},\ldots,\vec{W}^{x,\L})\},
\end{split}
\label{DLMx}
\end{equation} 
where
$\vec{F}_{t_i}^{x}=\textrm{diag}(\vec{F}_{t_i}^{x,1},\ldots,\vec{F}_{t_i}^{x,\L})$,
$\vec{\theta}_{t_i}^x=((\vec{\theta}_{t_i}^{x,1})^T,\ldots,(\vec{\theta}_{t_i}^{x,\L})^T)^T$
and the $3\L$-vector of spatial effects
$\vec{p}_{i}^x=((\vec{p}_{i}^{x,1})^T,\ldots,(\vec{p}_{i}^{x,\L})^T)^T$
is normally distributed with zero mean and covariance matrix
\[
\vec{K}^x= \begin{pmatrix} f^x(d_{11})\mathbb{I}_3 &  \hdots & f^x(d_{1\L})\mathbb{I}_3\\
\vdots & \ddots & \vdots\\
f^x(d_{\L1})\mathbb{I}_3 & \hdots & f^x(d_{\L\L})\mathbb{I}_3\end{pmatrix}.
\] 

\subsubsection{Additional harmonics}

Additional harmonics can be incorporated by using a Fourier form
structure \citep[see e.g.][]{West99,Petris09}. For ease of exposition,
we assume regularly spaced data at times $t_i=i-1,i=1,\ldots,n$. The
observation matrix in (\ref{obsx}) is defined to be the $1\times
(2q+1)$ matrix partitioned as $\vec{F}_{t_{i}}^{x,j}=(1, 0 | 1, 0 |
\ldots | 1)$ so that the state vector $\vec{\theta}_{t_i}^{x,j}$ is of
length $2q+1$ and satisfies a system equation of the form (\ref{sysx})
with system matrix
$\vec{G}^{x,j}_{t_i}=\textrm{diag}(\vec{H}_{1},\ldots,\vec{H}_{q},1)$,
where the $\vec{H}_{r}$ are harmonic matrices
\[
\vec{H}_{r} = 
\begin{pmatrix*}[r]
\cos\left(\pi  r/12\right) & \sin\left(\pi  r/12\right) \\
-\sin\left(\pi  r/12\right) & \cos\left(\pi  r/12\right)
\end{pmatrix*},\quad r=1,\ldots,q.
\]
The number of harmonics $q$ must be specified by the practitioner.
Note that for the full spatial temperature DLM, specifying $q$
harmonics will give $2\L(q+1)+6$ static parameters to be inferred.
Consequently, Fourier models with $q=1$ or~$2$ are typically used in
practice \citep{Petris09}. For the $q=1$ harmonic and the trivial case
of $\vec{W}^{x,j}=\vec{0}$, the observation equation of the Fourier
form DLM coincides with that the sinusoidal form in (\ref{obsx}) given
by
\[
X_{t_i}^j= \theta_{0,1}^{x,j}\cos\left(\pi t_i/12\right) + \theta_{0,2}^{x,j}\sin\left(\pi t_i/12\right) +\theta_{0,3}^{x,j}+v_i^{x,j}.
\] 
However, when $\vec{W}^{x,j}\neq\vec{0}$, the error structures differ due
to the use of the harmonic in the system equation of the Fourier form
DLM, and in the observation equation for the sinusoidal form DLM.  The
task of choosing between competing models is considered in Appendix
\ref{app:modelselec}.

\subsection{Spatial humidity DLM}
\label{sec:hdlm}

Due to the strong linear relationship between temperature and
humidity, we specify a conditional DLM for humidity by regressing on
temperature in the observation equation. For a particular location
$j$, the DLM takes the form
\begin{align*}
Y_{t_i}^j &= \vec{F}_{t_i}^{y,j}\vec{\theta}_{t_i}^{y,j}+v_{i}^{y,j},& v_{i}^{y,j}&\indep N\left(0,V^{y,j}\right)  \\  
\vec{\theta}_{t_i}^{y,j} &=\vec{\theta}_{t_{i-1}}^{y,j}+k_{i}\vec{w}_{i}^{y,j}+\vec{p}_i^{y,j},& \vec{w}_{i}^{y,j}&\indep N\{\vec{0},\textrm{diag}(\vec{W}^{y,j})\}
\end{align*}
where $\vec{F}_{t_i}^{y,j}=(X_{t_i}^j, 1)$,
$\vec{\theta}_{t_i}^{y,j}=(\theta_{t_i,1}^{y,j},\theta_{t_i,2}^{y,j})^T$
and $\vec{W}^{y,j}=(W_{1}^{y,j},W_{2}^{y,j})^T$. As in
Section~\ref{sec:tdlm}, we assign the components of the spatial error
process $\vec{p}_i^{y,j}=(p_{i,1}^{y,j},p_{i,2}^{y,j})^T$ independent
zero mean Gaussian process priors with covariance functions
\[
f_m^y(d_{jj'})=\textrm{Cov}(\theta_{t_i,m}^{y,j},\theta_{t_i,m}^{y,j'})
=\sigma_{y,m}^2\exp(-\psi_{y,m} d_{jj'}),\quad m=1,2.
\]
The spatial humidity DLM then takes the form
\begin{equation}
\begin{split}
\vec{Y}_{t_i} &= \vec{F}_{t_i}^{y}\vec{\theta}_{t_i}^{y}+\vec{v}_{i}^{y},\qquad \vec{v}_{i}^{y}\indep N\{\vec{0},\textrm{diag}(V^{y,1},\ldots,V^{y,\L})\}  \\  
\vec{\theta}_{t_i}^y &=\vec{\theta}_{t_{i-1}}^{y}+k_{i}\vec{w}_{i}^{x}+\vec{p}_i^y,\qquad \vec{w}_{i}^{y}\indep N\{\vec{0},\textrm{diag}(\vec{W}^{y,1},\ldots,\vec{W}^{y,\L})\}
\end{split}
\label{DLMy}
\end{equation} 
where
$\vec{F}_{t_i}^{y}=\textrm{diag}(\vec{F}_{t_i}^{y,1},\ldots,\vec{F}_{t_i}^{y,\L})$,
$\vec{\theta}_{t_i}^y=((\vec{\theta}_{t_i}^{y,1})^T,\ldots,(\vec{\theta}_{t_i}^{y,\L})^T)^T$
and the $2\L$-vector of spatial effects $\vec{p}_{i}^y$ is distributed
analogously to $\vec{p}_i^x$. Note that the joint model given by
(\ref{DLMx}) and (\ref{DLMy}) induces a marginal model for hourly
average humidity with the sinusoidal pattern observed in the data.
After integrating out $X_{t_i}^j$ in the observation equation for
$Y_{t_i}^j$, we obtain
\[
Y_{t_i}^j = \vec{F}_{t_i}^{x,j}\vec{\theta}_{t_i}^{x,j}\theta_{t_i,1}^{y,j}+\theta_{t_i,2}^{y,j}+v_{i}^{y,j}+\theta_{t_i,1}^{y,j}v_{i}^{x,j}
\]
which exhibits the same sinusoidal structure of (\ref{obsx}), albeit
with a different amplitude, phase and basal level. It is clear that
the joint model for $(X_{t_i}^j,Y_{t_i}^j)^T$ is not a DLM, as the
marginal humidity model depends on $\vec{\theta}_{t_i}^{x,j}$ and
$\vec{\theta}_{t_i}^{y,j}$ in a nonlinear way. Nevertheless, the
factorisation of the joint model as marginal and conditional DLMs can
be exploited when performing inference for the model parameters, and
this is the subject of the next section.

\section{Sequential Bayesian inference}\label{sec:bayes}

\subsection{Setup}\label{setup}
Fitting the model for temperature and humidity described in
Section~\ref{sec:dlm} to data is complicated by the fact that in
practice, sensor data is sometimes missing at one or more locations.
To deal with this scenario, we let $\vec{X}_{t_i}^{o}$ and
$\vec{Y}_{t_i}^{o}$ denote the observed temperature and humidity
processes at time~$t_i$.  We assume that if temperature is missing at
location $j$ at time $t_i$, then so is humidity (and vice-versa), as
is the case for our application. The observation model can then be
written as
\begin{equation}\label{obsmod}
\vec{X}_{t_i}^{o}=\vec{P}_{t_i}\vec{X}_{t_i},\qquad \vec{Y}_{t_i}^{o}=\vec{P}_{t_i}\vec{Y}_{t_i}
\end{equation} 
where the $n_i \times \L$ incidence matrix $\vec{P}_{t_i}$ determines
which components are observed at time $t_i$. For example, if we have
data streams from 5 different locations and data are missing at the
second and third location at time $t_i$, then the incidence matrix is
\[
\vec{P}_{t_i}=
\begin{pmatrix}
1& 0& 0& 0& 0 \\
0& 0& 0& 1& 0 \\
0& 0& 0& 0& 1 \\
\end{pmatrix}.
\]

Let $\vec{\phi}_x$ denote the flattened vector of
$V^{x,1},\ldots,V^{x,\L}$, $\vec{W}^{x,1},\ldots,\vec{W}^{x,\L}$,
$\vec{\sigma}_x$ and $\vec{\psi}_x$. Define $\vec{\phi}_y$ similarly.
Given observations $\vec{x}_{0:t_i}^{o}$ and $\vec{y}_{0:t_i}^{o}$ at
times $0=t_1<t_2<\ldots <t_i$, our primarily goal is sequential
exploration of the marginal posterior density
$\pi(\vec{\phi}_x,\vec{\phi}_y | \vec{x}_{0:t_i}^{o},
\vec{y}_{0:t_i}^{o})$. We assume that $\vec{\phi}_x$ and
$\vec{\phi}_y$ are independent \emph{a priori} with prior density
$\pi(\vec{\phi}_x,\vec{\phi}_y)=\pi(\vec{\phi}_x)\pi(\vec{\phi}_y)$.
Bayes' theorem gives the posterior density of interest as
\begin{align}
\pi(\vec{\phi}_x,\vec{\phi}_y | \vec{x}_{0:t_i}^{o}, \vec{y}_{0:t_i}^{o})&\propto \pi(\vec{\phi}_x)\pi(\vec{\phi}_y)\pi(\vec{x}_{0:t_i}^{o}, \vec{y}_{0:t_i}^{o}|\vec{\phi}_x,\vec{\phi}_y)\nonumber\\ 
&= \pi(\vec{\phi}_x)\pi(\vec{\phi}_y)\pi(\vec{x}_{0:t_i}^{o}|\vec{\phi}_x)\pi(\vec{y}_{0:t_i}^{o}|\vec{x}_{0:t_i}^{o}\vec{\phi}_y)\nonumber\\
&\propto  \pi(\vec{\phi}_x | \vec{x}_{0:t_i}^{o})\pi(\vec{\phi}_y | \vec{x}_{0:t_i}^{o},\vec{y}_{0:t_i}^{o})\label{postfact}
\end{align}
and so the parameter sets $\vec{\phi}_x$ and $\vec{\phi}_y$ are
independent \emph{a posteriori}. Moreover, we have that
\begin{equation}
\begin{split}
\pi(\vec{\phi}_x | \vec{x}_{0:t_i}^{o})                       &\propto\pi(\vec{\phi}_x|\vec{x}_{0:t_{i-1}}^{o})\pi(\vec{x}^o_{t_i}|\vec{x}^o_{0:t_{i-1}},\vec{\phi}_x)    \\  
\pi(\vec{\phi}_y | \vec{x}_{0:t_i}^{o},\vec{y}_{0:t_i}^{o}) &\propto\pi(\vec{\phi}_y|\vec{x}_{0:t_{i-1}},\vec{y}_{0:t_{i-1}}^{o})\pi(\vec{y}^o_{t_i}|x_{0:t_{i}},\vec{y}^o_{0:t_{i-1}},\vec{\phi}_y)
\end{split}
\label{margll}
\end{equation} 
where the observed data likelihood contributions
$\pi(\vec{x}^o_{t_i}|\vec{x}^o_{0:t_{i-1}},\vec{\phi}_x)$ and
$\pi(\vec{y}^o_{t_i}|\vec{x}^o_{0:t_{i}},\vec{y}^o_{0:t_{i-1}},\vec{\phi}_y)$
can be calculated using a forward filter \citep{West99}. Details of
this calculation can be found in Appendix~\ref{sec:ff}.

\subsection{Iterated batch importance sampling}
\label{subsec:IBIS}

Although the parameter posterior is intractable, the form of
(\ref{margll}) suggests a sequential importance sampling scheme that
repeatedly reweights a set of $N$ parameter samples (known as
`particles' in this context) by the observed data likelihood
contributions. This approach is used in the iterated batch importance
sampling (IBIS) algorithm of \cite{Chopin02}, together with MCMC steps
for rejuvenating parameter samples in order to circumvent particle
degeneracy. Given the factorisation of the posterior in
(\ref{postfact}), in what follows we focus on recursive sampling from
$\pi(\vec{\phi}_x | \vec{x}_{0:t_i}^{o})$ and note that the steps for
sampling from $\pi(\vec{\phi}_y |
\vec{x}_{0:t_i}^{o},\vec{y}_{0:t_i}^{o})$ are similar.

Suppose that a weighted sample
$\{\vec{\phi}_x^{(k)},\omega_{t_i}^{(k)}\}_{k=1}^{N}$ from
$\pi(\vec{\phi}_x | \vec{x}_{0:t_i}^{o})$ is available. The IBIS
algorithm involves two steps: an incremental weighting step and a
rejuvenation (resample-move) step. In the incremental weight step, the
weight is updated for each particle through the observed data
likelihood contribution of the current observation, i.e.
$\omega_{t_i}^{(k)}\propto\omega_{t_{i-1}}^{(k)}\pi(\vec{x}^o_{t_i}|\vec{x}^o_{0:t_{i-1}},\vec{\phi}_x^{(k)})$.
Note that the calculation of the observed data likelihood increment
(as given by the forward filter in Appendix~\ref{sec:ff}) requires the
posterior summaries
$\vec{m}_{t_{i-1}}(\vec{\phi}_x^{(k)})=\vec{m}_{t_{i-1}}^{(k)}$ and
$\vec{C}_{t_{i-1}}(\vec{\phi}_x^{(k)})=\vec{C}_{t_{i-1}}^{(k)}$ of
$\pi(\vec{\theta}_{t_i}^x|\vec{x}^o_{0:t_{i-1}},\vec{\phi}^x)$.

Simply updating the incremental weights over the time will lead to
particle degeneracy. To bypass this problem, the IBIS scheme uses a
resample-move step \citep[see e.g.][]{gilks2001} that firstly
resamples parameter particles (e.g. by drawing indices from a
multinomial $\mathcal{M}(\omega^{1:N})$ distribution) and then moves
each parameter sample through a Metropolis-Hastings kernel which
leaves the target posterior invariant. The resample-move step is only
used if some degeneracy criterion is fulfilled. Typically, at each
time $t_i$, the effective sample size (ESS) is computed as
\[
\textrm{ESS}=1\big / \,\,{\sum_{k=1}^{N}(\omega_{t_i}^{(k)})^2}
\] 
and the resample-move step is triggered if $\textrm{ESS}<\delta N$ for
$\delta \in(0,1)$ and a standard choice is $\delta=0.5$. As the
parameters must be strictly positive, we take a proposal density
\[
q(\vec{\phi}_x^*|\vec{\phi}_x)=\log N\left\{\vec{\phi}_x^*; \log\vec{\phi}_x,\gamma Var(\log\vec{\phi}_x|\vec{x}_{0:t_{i}}^o)\right\}
\]
where $\log N(\cdot;\vec{m},\vec{V})$ denotes the density associated
with the exponential of a $N(\vec{m},\vec{V})$ random variable. We use
the standard rule of thumb of \cite{Roberts97} and \cite{Roberts01} by
taking the scaling parameter $\gamma= 2.38^2/n_{par}$, where $n_{par}$
is the number of parameters. The full IBIS scheme is given by
Algorithm~\ref{IBIS}.

\begin{algorithm}[t]
\caption{IBIS scheme}\label{IBIS}
\begin{enumerate}
\item Initialisation. For $k=1,\ldots,N$ sample
  $\vec{\phi}_x^{(k)}\sim \pi(\cdot)$ and set
  $\tilde{\omega}_{0}^{(k)}=\pi(\vec{x}_0^o|\vec{\phi}_x^{(k)})$ using
  iteration $i=1$ of the forward filter. Store $\vec{m}_{t_1}^{(k)}$
  and $\vec{C}_{t_1}^{(k)}$. 
\item[] For $i=2,\ldots ,n$:
\item Sequential importance sampling. For $k=1,\ldots,N$:
\begin{itemize}
\item[(a)] Perform iteration $i$ of the forward filter to obtain $\pi(\vec{x}_{t_i}^o|\vec{x}_{0:t_{i-1}}^o,\vec{\phi}_x^{(k)})$, $\vec{m}_{t_i}^{(k)}$ and $\vec{C}_{t_i}^{(k)}$. Note the convention 
that $\pi(\vec{x}_0^o|\vec{\phi}_x^{(k)})=\pi(\vec{x}_{0}^o|\vec{x}_{0:t_1}^o,\vec{\phi}^{(k)}_x)$. 
\item[(b)] Update and normalise the importance weights using
\[
\tilde{\omega}_{t_i}^{(k)}=\tilde{\omega}_{t_{i-1}}^{(k)}\pi(\vec{x}_{t_i}^o|\vec{x}_{0:t_{i-1}}^o,\vec{\phi}_x^{(k)}), \qquad 
\omega_{t_{i}}^{(k)}=\frac{\tilde{\omega}_{t_i}^{(k)}}{\sum_{j=1}^{N}\tilde{\omega}_{t_{i}}^{(j)}}
\]
\item[(c)] Update the observed data likelihood using
\[
\pi(\vec{x}_{0:t_i}^o|\vec{\phi}_x^{(k)})=\pi(\vec{x}_{0:t_{i-1}}^o|\vec{\phi}_x^{(k)})\pi(\vec{x}_{t_i}^o|\vec{x}_{0:t_{i-1}}^o,\vec{\phi}_x^{(k)}).
\]
\end{itemize}
\item If $\textrm{ESS}< \delta N$ resample and move as follows. For $k=1,\ldots,N$:
\begin{itemize}
\item[(a)] Sample indices $a_{k}\sim \mathcal{M}\big(\omega^{1:N}\big)$ 
and set $\{\vec{\phi}_x^{(k)},\tilde{\omega}_{t_i}^{(k)}\}:=\{\vec{\phi}_x^{(a_{k})},1\}$, $\pi(\vec{x}_{0:t_i}^o|\vec{\phi}_x^{(k)}):=\pi(\vec{x}_{0:t_i}^o|\vec{\phi}_x^{(a_k)})$, 
$\vec{m}_{t_i}^{(k)}:=\vec{m}_{t_i}^{(a_k)}$ and $\vec{C}_{t_i}^{(k)}:=\vec{C}_{t_i}^{(a_k)}$.
\item[(b)] Propose $\vec{\phi}_x^*\sim q(\cdot|\vec{\phi}_x^{(k)})$. Perform iterations $1,\ldots,i$ 
of the forward filter to obtain $\pi(\vec{x}_{0:t_i}^o|\vec{\phi}_x^{*})$. With probability
\[
\textrm{min}\left\{1, \frac{\pi(\vec{\phi}_x^*)\pi(\vec{x}_{0:t_i}^o|\vec{\phi}_x^{*})}{\pi(\vec{\phi}_x^{(k)})\pi(\vec{x}_{0:t_i}^o|\vec{\phi}_x^{(k)})}\times 
\frac{q(\vec{\phi}_x^{(k)}|\vec{\phi}_x^*)}{q(\vec{\phi}_x^*|\vec{\phi}_x^{(k)})}\right\}
\]
put $\vec{\phi}_x^{(k)}:=\vec{\phi}_x^*$, $\pi(\vec{x}_{0:t_i}^o|\vec{\phi}_x^{(k)}):=\pi(\vec{x}_{0:t_i}^o|\vec{\phi}_x^{*})$, $\vec{m}_{t_i}^{(k)}:=\vec{m}_{t_i}^*$ and $\vec{C}_{t_i}^{(k)}:=\vec{C}_{t_i}^*$.
\end{itemize}
\end{enumerate}
\end{algorithm}

Finally, we note that it is straightforward to estimate the evidence  
\[
\pi(\vec{x}_{0:t_{n}}^o)=\prod_{i=1}^{n}\pi(\vec{x}_{t_i}^o|\vec{x}_{0:t_{i-1}}^o) 
\]
using the output of the IBIS scheme, at virtually no additional computational cost. 
Each factor $L_{t_i}=\pi(\vec{x}_{t_i}^o|\vec{x}_{0:t_{i-1}}^o)$ in the product above is estimated by
\begin{equation}\label{ev}
L_{t_{1}}=\sum_{k=1}^{N}\frac{1}{N}\pi(\vec{x}_0^o|\vec{\phi}_x^{(k)}),\qquad L_{t_i}=\sum_{k=1}^{N}\omega_{t_{i-1}}^{(k)} \pi(\vec{x}_{t_i}^o|\vec{x}_{0:t_{i-1}}^o,\vec{\phi}_x^{(k)}), \quad i=2,\ldots,n.
\end{equation}    

\subsection{Online IBIS}
The main computational bottleneck of IBIS is the resample-move step.
If this step is triggered at time $t_i$, then the observed data
likelihood $\pi(\vec{x}_{0:t_i}^o|\vec{\phi}_x^{*})$ must be
calculated for each proposed particle $\vec{\phi}_x^{*}$.
Consequently, the computational cost grows with $t_i$, precluding the
use of IBIS as an online scheme. To bound the computational cost of
assimilating a single observation, we modify the resample-move step by
basing the observed data likelihood on an observation window whose
time length is chosen to balance accuracy and computational
efficiency.

We follow a similar approach introduced by \cite{Moral17} and define a
sequence of windows with equal widths, say $T$, over the observation
period. First the observation period is divided into $b$ windows and denote by $\vec{x}^o_{t_i^s}$ the $i$th
observation in window $s\in
\{1, \ldots, b\}$, for $i = 1,\ldots, n_s$. The observation
times satisfy $t_i^{s} \in ( (s-1)T, sT]$ when $s = 1,\ldots, b-1$ and
$t_i^s \in ( (b-1)T, t^b_{n_b}]$ when $s = b$. The standard IBIS
scheme is run over the first window. For windows $s=2,\ldots, b$, the
resample-move step targets
\begin{equation}\label{approxpost}
\tilde{\pi}(\vec{\phi}_x|\vec{x}^o_{0:t_i^s})\propto \tilde{\pi}(\vec{\phi}_x|\vec{x}_{0:(s-1)T})\pi(\vec{x}^o_{t_{1}^s:t_i^s}|\vec{x}^o_{0:(s-1)T},\vec{\phi}_x)
\end{equation}
where
\[
\tilde{\pi}(\vec{\phi}_x|\vec{x}_{0:(s-1)T})=\frac{1}{N}\sum_{k=1}^{N}\log N(\vec{\phi}_x; \log\vec{\phi}_x^{(k)}, h_s^2)
\]
is a kernel density estimate (KDE) of
$\pi(\vec{\phi}_x|\vec{x}_{0:(s-1)T})$ and the bandwidth $h_s^2$ can
be calculated using, for example, Silverman's rule of
thumb~\citep{Silverman86} as
\[
h_s^2 = 1.06^2N^{-2/5}\widehat{Var}(\vec{\phi}_x^{(1:N)} | \vec{x}^o_{0:(s-1)T}).
\]
Thus in order to evaluate (\ref{approxpost}), we need only evaluate
the observed data likelihood contribution from the beginning of the
current window until the current time.  Furthermore, by taking the
proposal density to be
$q(\vec{\phi}_x^*|\vec{\phi}_x)=\tilde{\pi}(\vec{\phi}_x^*|\vec{x}_{0:(s-1)T})$,
the kernel density estimate need not be evaluated in the MH acceptance
ratio. The choice of the window width has a direct influence on
computational efficiency and posterior accuracy. A simulation study
comparing IBIS and online IBIS for different window lengths is given
in Section~\ref{subsubsec:onlineIBIS}. The online IBIS scheme is
summarised by Algorithm~\ref{onlineIBIS}.

\begin{algorithm}[t!]
\caption{Online IBIS scheme}\label{onlineIBIS}
\begin{enumerate}
\item Initialisation. Divide the observed period into $b$ windows, $s \in \{1, \ldots, b\}$. Denote by $t_i^{s}$ the $i$th observation time in window $s$, 
$i = 1,\ldots, n_s$. For $s = 1$, implement the IBIS scheme (Algorithm~\ref{IBIS}). For $s = 2, \ldots, b$ and $i = 1,\ldots, n_s$:
\item Sequential importance sampling. For $k=1,\ldots,N$:
\begin{itemize}
\item[(a)] Perform iteration $i$ (corresponding to time $t_i^s$) of the forward filter to obtain 
$\pi(\vec{x}_{t^s_i}^o|\vec{x}_{0:t^s_{i-1}}^o,\vec{\phi}_x^{(k)})$,  $\vec{m}_{t_i^s}^{(k)}$ and $\vec{C}_{t_i^s}^{(k)}$.
\item[(b)] Update and normalise the importance weights using
\[
\tilde{\omega}_{t^s_i}^{(k)}=\tilde{\omega}_{t^s_{i-1}}^{(k)}\pi(\vec{x}_{t^s_i}^o|\vec{x}_{0:t^s_{i-1}}^o,\vec{\phi}_x^{(k)}), \qquad 
\omega_{t^s_{i}}^{(k)}=\frac{\tilde{\omega}_{t^s_i}^{(k)}}{\sum_{z=1}^{N}\tilde{\omega}_{t^s_{i}}^{(z)}}
\]
\item[(c)] Update the observed data likelihood contribution in the current window using
\[
\pi(\vec{x}_{t_1^s:t_i^s}^o|\vec{x}^o_{0:(s-1)T},\vec{\phi}_x^{(k)})=\pi(\vec{x}_{t_1^s:t^s_{i-1}}^o|\vec{x}^o_{0:(s-1)T},\vec{\phi}_x^{(k)})\pi(\vec{x}_{t^s_i}^o|\vec{x}_{0:t^s_{i-1}}^o,\vec{\phi}_x^{(k)}),
\]
with the convention that $\pi(\vec{x}_{t_1^s:t_i^s}^o|\vec{x}^o_{0:(s-1)T},\vec{\phi}_x^{(k)})=\pi(\vec{x}_{t_1^s}|\vec{x}^o_{0:(s-1)T},\vec{\phi}_x^{(k)})$ for $i=1$.
\end{itemize}
\item If $\textrm{ESS}< \delta N$ resample and move. For $k=1,\ldots,N$:
\begin{itemize}
\item[(a)] Sample indices $a_{k}\sim \mathcal{M}\big(\omega^{1:N}\big)$ 
and set $\{\vec{\phi}_x^{(k)},\tilde{\omega}_{t^s_i}^{(k)}\}:=\{\vec{\phi}_x^{(a_{k})},1\}$, 
$\vec{m}_{t_i^s}^{(k)}:=\vec{m}_{t_i^s}^{(a_k)}$, $\vec{C}_{t_i^s}^{(k)}:=\vec{C}_{t_i^s}^{(a_k)}$ and
$\pi(\vec{x}_{t_1^s:t_i^s}^o|\vec{x}^o_{0:(s-1)T},\vec{\phi}_x^{(k)}):=\pi(\vec{x}_{t_1^s:t_i^s}^o|\vec{x}^o_{0:(s-1)T},\vec{\phi}_x^{(a_k)})$.
\item[(b)] Propose $\bm{\phi}_x^* \sim\log N(\log\bm{\phi}_x^{(k)}, h_s^2)$. Using $\vec{m}_{(s-1)T}^*=\vec{m}_{(s-1)T}^{(k)}$ and $\vec{C}_{(s-1)T}^*=\vec{C}_{(s-1)T}^{(k)}$, 
perform iterations $1,\ldots,i$ (corresponding to times $t^s_1,\ldots,t^s_i$) 
of the forward filter to obtain $\pi(\vec{x}_{t^s_1:t^s_i}^o|\vec{x}^o_{0:(s-1)T}, \vec{\phi}_x^{*})$. With probability
\[
\textrm{min}\left\{1, \frac{\pi(\vec{x}_{t^s_1:t^s_i}^o|\vec{x}^o_{0:(s-1)T},\vec{\phi}_x^{*})}{\pi(\vec{x}_{t^s_1:t^s_i}^o|\vec{x}^o_{0:(s-1)T},\vec{\phi}_x^{(k)})} 
\right\}
\]
put $\vec{\phi}_x^{(k)}:=\vec{\phi}_x^*$, $\pi(\vec{x}_{t^s_1:t^s_i}^o|\vec{x}^o_{0:(s-1)T},\vec{\phi}_x^{(k)}):=\pi(\vec{x}_{t^s_1:t^s_i}^o|\vec{x}^o_{0:(s-1)T},\vec{\phi}_x^{*})$, 
$\vec{m}_{t_i^s}^{(k)}:=\vec{m}_{t_i^s}^*$ and $\vec{C}_{t_i^s}^{(k)}:=\vec{C}_{t_i^s}^*$.
\end{itemize}
\end{enumerate}
\end{algorithm}

\subsection{Parallelising the algorithm}
The incremental weighting steps are readily parallelised in an SMC
scheme. Additionally, for IBIS the move step can be performed
independently for each particle. However, commonly used resampling
schemes, such as the multinomial approach considered here, involve a
collective operation (summing the weights) precluding obvious
parallelisation of the full IBIS scheme. \cite{Hendeby10} and
\cite{Gong12} describe a forward adder tree method which parallelises
the calculation of the cumulative weight. \cite{Murray16} suggest
parallel Metropolis resampling and rejection resampling schemes to
mitigate numerical instabilities of summing the weights for a large
number of particles. However, these methods still require information
exchange and global operations and they are designed mainly for use on
GPU shared memory systems.

Distributed memory systems are naturally amenable to heavy
parallelised jobs, where trunks of jobs are allocated and processed
over multiple cores in different processors. In this context, a number
of parallel resampling methods have been discussed in the literature;
see, for example, \cite{Brun02} and \cite{Bolic04,Bolic05}. We follow
the local resampling method \citep{Brun02} by partitioning particles
into disjoint subsets, within which resampling is performed.  The
algorithm proceeds by first calculating a local ESS for each subset of
particles. If a local ESS is less than a threshold, then the
rejuvenation step is triggered locally. The innovation variance for
the MH proposal in the move step is also calculated locally based on
the individual particle subset. To mitigate load-balance problems that
can occur when the resample-move step is executed for some subsets but
not others, we also carry out a rejuvenation step at regular time
points, e.g. every 20 time points. This approach naturally fits within
the distributed memory architecture and allows full parallelisation of
the IBIS scheme. In principle, this approach should significantly
improve computational efficiency of the inference scheme, as there is
no need for task communication. However, in practice the number of
informative particles may reduce significantly in some subsets as the
algorithm runs. This in turn results in the rejuvenation step being
executed more frequently.  Therefore, a trade-off has to be considered
carefully between the number of particle subsets and the number of
particles in each subset.  Section~\ref{subsubsec:paraRe} describes a
simulation study comparing a standard serial implementation with a
fully parallelised version (with local resampling).

\section{Simulation study}

In order to assess the performance of the proposed online IBIS scheme
and the effect of local resampling, we looked at results from
synthetic data generated from the marginal model in (\ref{DLMx}). We
consider 2~spatial locations (giving 14 parameters in total) and
simulated $n=1300$ observations at each location. The true parameter
values used to produce the synthetic data are $W^j_k=0.01$,
$V^j=\sigma_k^2=1$ and $\psi_k=0.01$ for $j=1,2$ and $k=1,2,3$, and
these values are shown in Figure~\ref{paraRe-serialRe}. As this is a
data-rich scenario, we assumed very weak independent inverse Gamma
$IG(1,0.01)$ prior distributions for all these parameter components,
but truncated them above at 10 as values in excess of~10 are far from
plausible. We also took the prior distribution for the initial system
state as $\bm{\theta}_0\sim N(\vec{m},\,\vec{C})$, where
$\vec{m}=(0,0,17,0,0,17)^T$ and $\vec{C}=\mathbb{I}_6$. We used $10^7$
particles and an ESS threshold of $\delta=0.5$ for triggering the
resample-move step. All computer code was written in C and executed on
a high performance cluster with Intel Xeon E5-2699 v4 processors (2.2
GHz, 55 MB cache), where each processor has 22 cores (2.9 GB CPU
memory per
core). 

\subsection{Comparison of full IBIS with serial resampling and parallelised local resampling}
\label{subsubsec:paraRe}
\begin{figure}[p]
\centering
\includegraphics[width=1\textwidth,height=0.9\textheight]{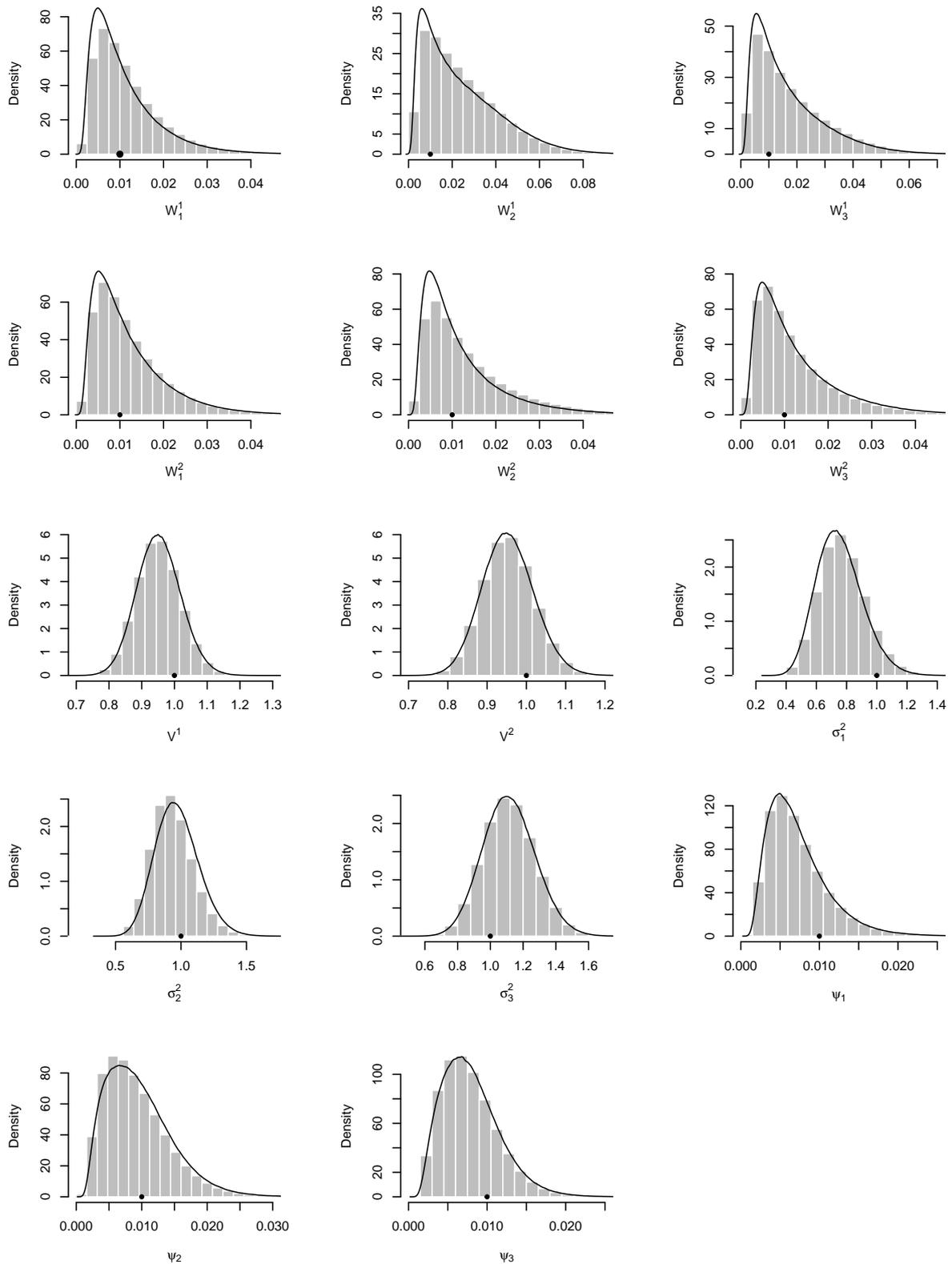}
\caption{Marginal parameter posterior densities obtained from the
  output of the full IBIS scheme with a standard serial resampling
  step (histograms) and a parallelised local resampling step (------).
  The true parameter values are shown as solid circles.}
\label{paraRe-serialRe}
\end{figure}

We consider first two parallelised implementations of the full IBIS
scheme: (i) weighting and move steps are performed in parallel over 22
cores through a shared memory system (within one processor) with the
resampling step performed in serial; (ii) particles are divided over
200 cores and local resampling is used.  Figure~\ref{paraRe-serialRe}
shows the parameter marginal posterior densities obtained by using
method 1 (IBIS with serial resampling) and method~2 (IBIS with
parallelised local resampling). It is clear that both approaches give
posterior output consistent with the true values (used to simulate the
data). Moreover, the posterior densities from the fully parallelised
method~2 match up well with those from the exact (simulation based)
method 1. However the run time for method~1 (IBIS with serial
resampling) is around 23 hours whereas that for method~2 (IBIS with
parallelised local resampling) is around 4 hours, a speed-up of around
a factor of 6.

\subsection{Comparison of full IBIS and online IBIS}
\label{subsubsec:onlineIBIS}
\begin{figure}[p]
\centering
\includegraphics[width=1\textwidth,height=0.9\textheight]{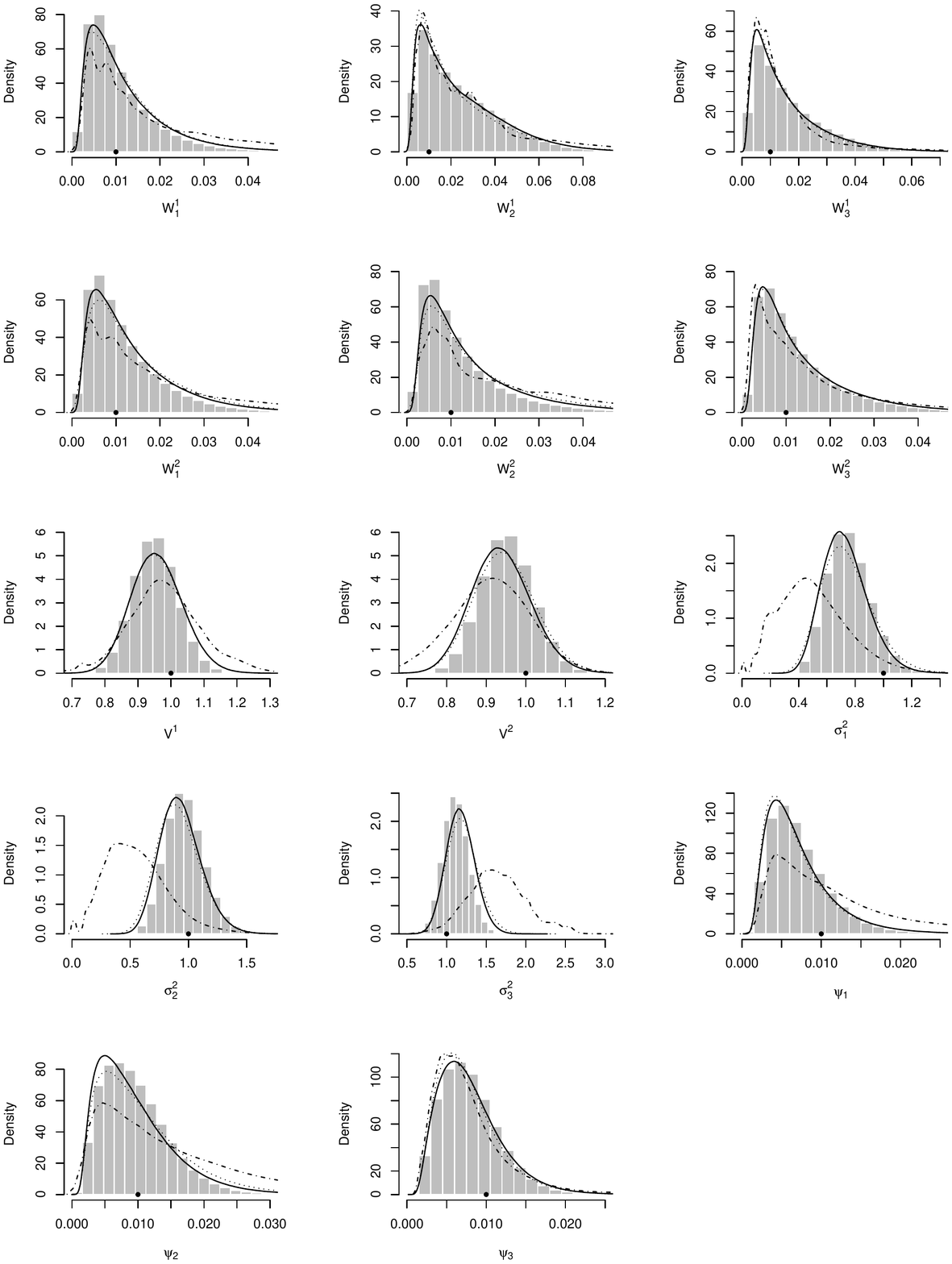}
\caption{Marginal parameter posterior densities obtained from the
  output of the full IBIS scheme (histograms) and the online IBIS
  scheme with window widths $T = 100$ ($- \cdot - \cdot -$), $T = 300$
  ($\cdots \cdots$) and $T = 500$ (------). The true parameter values
  are are shown as solid circles.}
\label{onlineIBISplot}
\end{figure}

We now compare the full IBIS scheme with online IBIS and in both
schemes we use the parallelised local resampling method. For online
IBIS, we consider three widths for the fixed window: $T = 100$, $300$
and $500$. Figure \ref{onlineIBISplot} shows the output of the
marginal posterior densities from the online IBIS scheme for each
window size, together with the densities from the full IBIS scheme. As
expected, as the larger window increases, so does posterior accuracy.
The marginal posteriors from online IBIS using $T=300$ and $T=500$
almost overlay those from full IBIS. However, there are noticeable
differences when using $T=100$. In terms of computational efficiency,
online IBIS with both $T=300$ and $T=500$ take roughly 2~CPU hours,
that with $T=100$ takes approximately 1~CPU hour.  Consequently, for
this example, online IBIS with $T=300$ and local parallel resampling
gives an overall reduction in computational cost of around a factor of
12 compared to full IBIS with serial resampling.

\section{Application}
\label{app}
In this section we analyse the data on hourly average temperature and
humidity values introduced in Section~\ref{sec:data}.  Recall that
these data are measurements recorded during the period 8th July 2017
to 31st December 2017 and that the observations are irregularly spaced
due to network and sensor failures. We take independent inverse Gamma
$IG(1,0.01)$ prior distributions, truncated above at~10, for all the
static parameters in both temperature and humidity DLMs.  To
incorporate our prior belief that the underlying system should be
smoother than the observation process, we also impose the constraint
that at each location $j=1,\ldots,5$, $W^{x,j}_i < V^{x,j}$
($i=1,2,3$) and $W^{y,j}_k < V^{y,j}$ ($k=1,2$).  We ran the online
IBIS scheme with $N=10^7$ particles, fully parallelised (with local
resampling) over 200 cores using an ESS threshold of $\delta =0.5$.
Regular particle rejuvenation steps were set up for the process at
every 20 time points, and the resample-move step was executed in any
batch whose ESS fell below half the number of particles (in the
batch). Finally, to balance accuracy and computational efficiency, we
used a window width of $T=1500$, and this gave a run time of
approximately 9.5 days.

\subsection{Inference results}\label{sec:results}
\begin{figure}[t]
\centering
\includegraphics[width=1\textwidth,height=0.54\textheight]{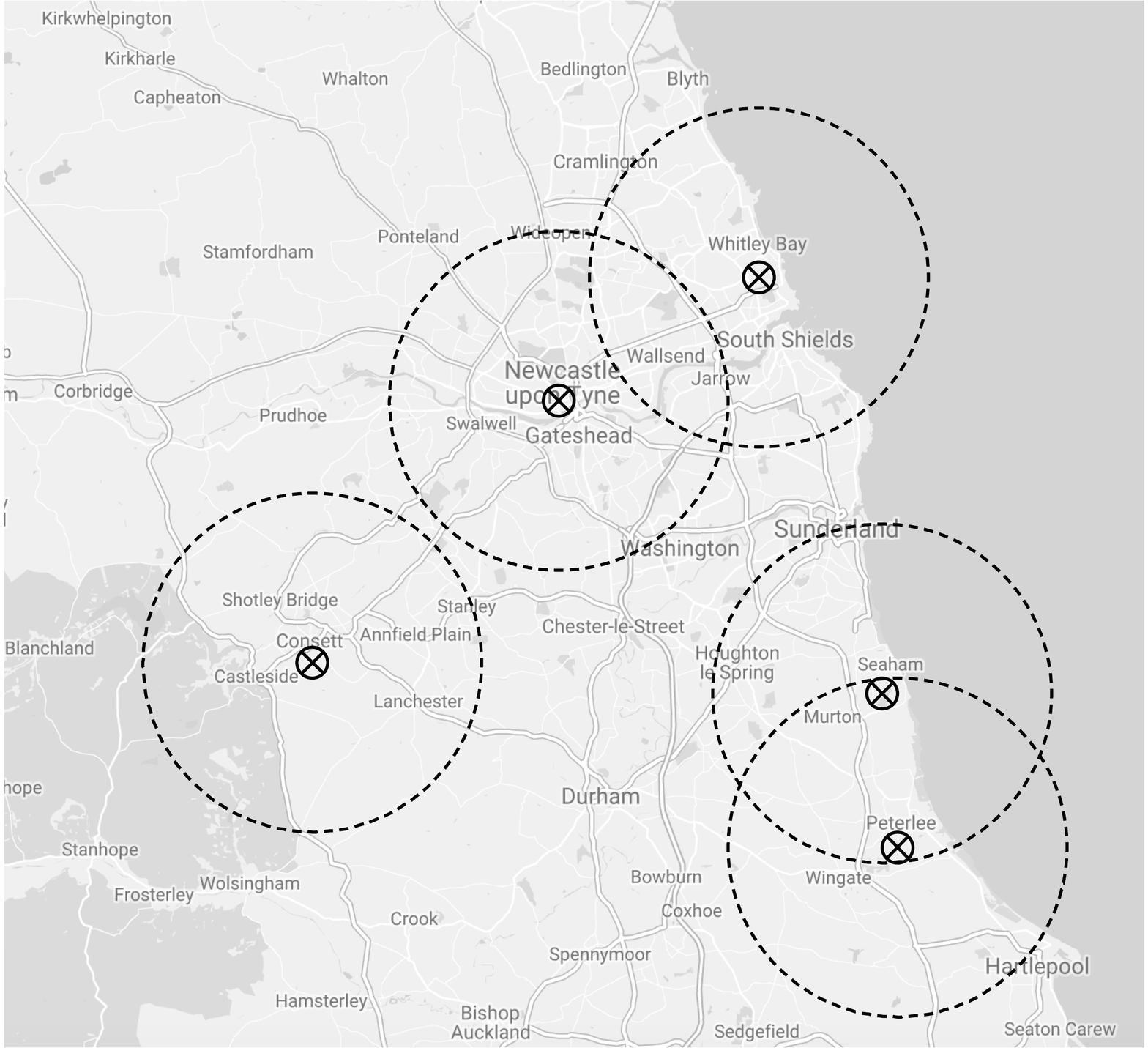}
\caption{Map showing site locations and a 10\,km radius from each
  site, within which the spatial correlation for temperature is at
  least 0.76, and for humidity, is at least 0.64.}
\label{map}
\end{figure}

Table~\ref{postandsd} shows the marginal posterior medians and
quantile-based 95\% credible intervals for the static parameters in
the joint temperature and humidity model. These summaries were
obtained from output of the online IBIS scheme. Inspection of the
posterior medians for the system variances (governing both temperature
and humidity models) reveals that these components are larger at
location~1 (Newcastle) than at the other locations. This is perhaps
not surprising given that location~1 has the largest fraction of
missing data (see Table~\ref{missingdata}). Also sampled posterior
values of the observation variance components $V^{x,j}$ and $V^{y,j}$
are generally very much larger at location~2 (Seaham), and this too is
consistent with the simple data summaries in Table~\ref{missingdata}
-- Seaham is the least spatially consistent location in terms of
median temperature and humidity. Variation across sites is accounted
for by the elements of~$\vec{\sigma}^2$. The relatively large values
of $\sigma^2_{x,3}$ and $\sigma^2_{y,2}$ suggest that there is some
spatial inconsistency in the dynamically varying mean level components
$\theta_{t_i,3}^{x,j}$ and $\theta_{t_i,2}^{y,j}$. Spatial consistency
of these mean level components can be assessed further by noting that
\[
\textrm{Cor}(\theta_{t_i,3}^{x,j},\theta_{t_i,3}^{x,j'})
=\exp(-\psi_{x,3} d_{jj'}),\qquad 
\textrm{Cor}(\theta_{t_i,2}^{y,j},\theta_{t_i,2}^{y,j'})
=\exp(-\psi_{y,2} d_{jj'}).
\]
Hence, fixing $\psi_{x,3}$ and $\psi_{y,2}$ at their posterior medians
gives a simple linear relationship between distance and log
correlation. For example, within a 10km radius from each location,
there is a spatial correlation of at least 0.76 for temperature and
0.64 for humidity. These areas are displayed in Figure~\ref{map}. We
note that it is not surprising that spatial correlation for humidity
is lower than that for temperature, as the humidity records are also
easily influenced by other factors, such as urban structure and
distance from the sea, in addition to temperature.

\begin{table}[t!]
\centering
\begin{tabular}{cccc|cccc}
\toprule
\multicolumn{4}{c|}{Temperature} & \multicolumn{4}{c}{Humidity} \\ \hline
$\vec{\phi}_x$ & Median   & 2.5\%  & 97.5\%  & $\vec{\phi}_y$ & Median   & 2.5\%   & 97.5\%   \\ \toprule
$W_1^{x,1}$ & 0.0050 & 0.0011 & 0.0110 & $W_1^{y,1}$ & 0.0156 & 0.0118 & 0.0208 \\
$W_2^{x,1}$ & 0.0056 & 0.0013 & 0.0114 & $W_2^{y,1}$ & 0.0074 & 0.0019 & 0.0183 \\
$W_3^{x,1}$ & 0.0053 & 0.0014 & 0.0116 & $W_1^{y,2}$ & 0.0071 & 0.0049 & 0.0102 \\
$W_1^{x,2}$ & 0.0026 & 0.0008 & 0.0089 & $W_2^{y,2}$ & 0.0072 & 0.0018 & 0.0183 \\
$W_2^{x,2}$ & 0.0031 & 0.0008 & 0.0095 & $W_1^{y,3}$ & 0.0024 & 0.0014 & 0.0038 \\
$W_3^{x,2}$ & 0.0039 & 0.0009 & 0.0096 & $W_2^{y,3}$ & 0.0048 & 0.0015 & 0.0144 \\
$W_1^{x,3}$ & 0.0021 & 0.0006 & 0.0082 & $W_1^{y,4}$ & 0.0032 & 0.0017 & 0.0054 \\
$W_2^{x,3}$ & 0.0023 & 0.0006 & 0.0075 & $W_2^{y,4}$ & 0.0050 & 0.0016 & 0.0156 \\
$W_3^{x,3}$ & 0.0021 & 0.0006 & 0.0083 & $W_1^{y,5}$ & 0.0020 & 0.0010 & 0.0035 \\
$W_1^{x,4}$ & 0.0027 & 0.0007 & 0.0083 & $W_2^{y,5}$ & 0.0049 & 0.0016 & 0.0148 \\
$W_2^{x,4}$ & 0.0032 & 0.0007 & 0.0095 & $V^{y,1}$ & 0.0265 & 0.0147 & 0.0826 \\
$W_3^{x,4}$ & 0.0036 & 0.0009 & 0.0102 & $V^{y,2}$ & 0.4520 & 0.3362 & 0.5822 \\
$W_1^{x,5}$ & 0.0042 & 0.0008 & 0.0103 & $V^{y,3}$ & 0.0201 & 0.0137 & 0.0382 \\
$W_2^{x,5}$ & 0.0026 & 0.0007 & 0.0089 & $V^{y,4}$ & 0.0199 & 0.0137 & 0.0383 \\
$W_3^{x,5}$ & 0.0038 & 0.0007 & 0.0092 & $V^{y,5}$ & 0.0190 & 0.0134 & 0.0331 \\
$V^{x,1}$ & 0.0089 & 0.0047 & 0.0173 & $\sigma^2_{y,1}$ & 0.0257 & 0.0209 & 0.0315 \\
$V^{x,2}$ & 0.0230 & 0.0110 & 0.0419 & $\sigma^2_{y,2}$ & 1.6054 & 1.4961 & 1.7228 \\
$V^{x,3}$ & 0.0078 & 0.0044 & 0.0138 & $\psi_{y,1}$ & 0.0016 & 0.0008 & 0.0029 \\
$V^{x,4}$ & 0.0088 & 0.0049 & 0.0251 & $\psi_{y,2}$ & 0.0447 & 0.0388 & 0.0511 \\
$V^{x,5}$ & 0.0164 & 0.0061 & 0.0380 &    &          &         &    \\
$\sigma^2_{x,1}$ & 0.0423 & 0.0105 & 0.1611 &    &          &         &    \\
$\sigma^2_{x,2}$ & 0.0627 & 0.0250 & 0.1672 &    &          &         &    \\
$\sigma^2_{x,3}$ & 0.2310 & 0.0837 & 0.2706 &    &          &         &    \\
$\psi_{x,1}$ & 0.0014 & 0.0004 & 0.0496 &    &          &         &    \\
$\psi_{x,2}$ & 0.0013 & 0.0004 & 0.0606 &    &          &         &    \\
$\psi_{x,3}$ & 0.0274 & 0.0011 & 0.0354 &    &          &         &    \\ \toprule
\end{tabular}
\caption{Marginal parameter posterior medians and quantile-based 95\% credible intervals obtained from the output of the online IBIS scheme.}
\label{postandsd}
\end{table}

\subsection{Predictive performance}
\label{sec:predperf}
\begin{figure}[p]
\centering
\includegraphics[width=1\textwidth,height=0.8\textheight]{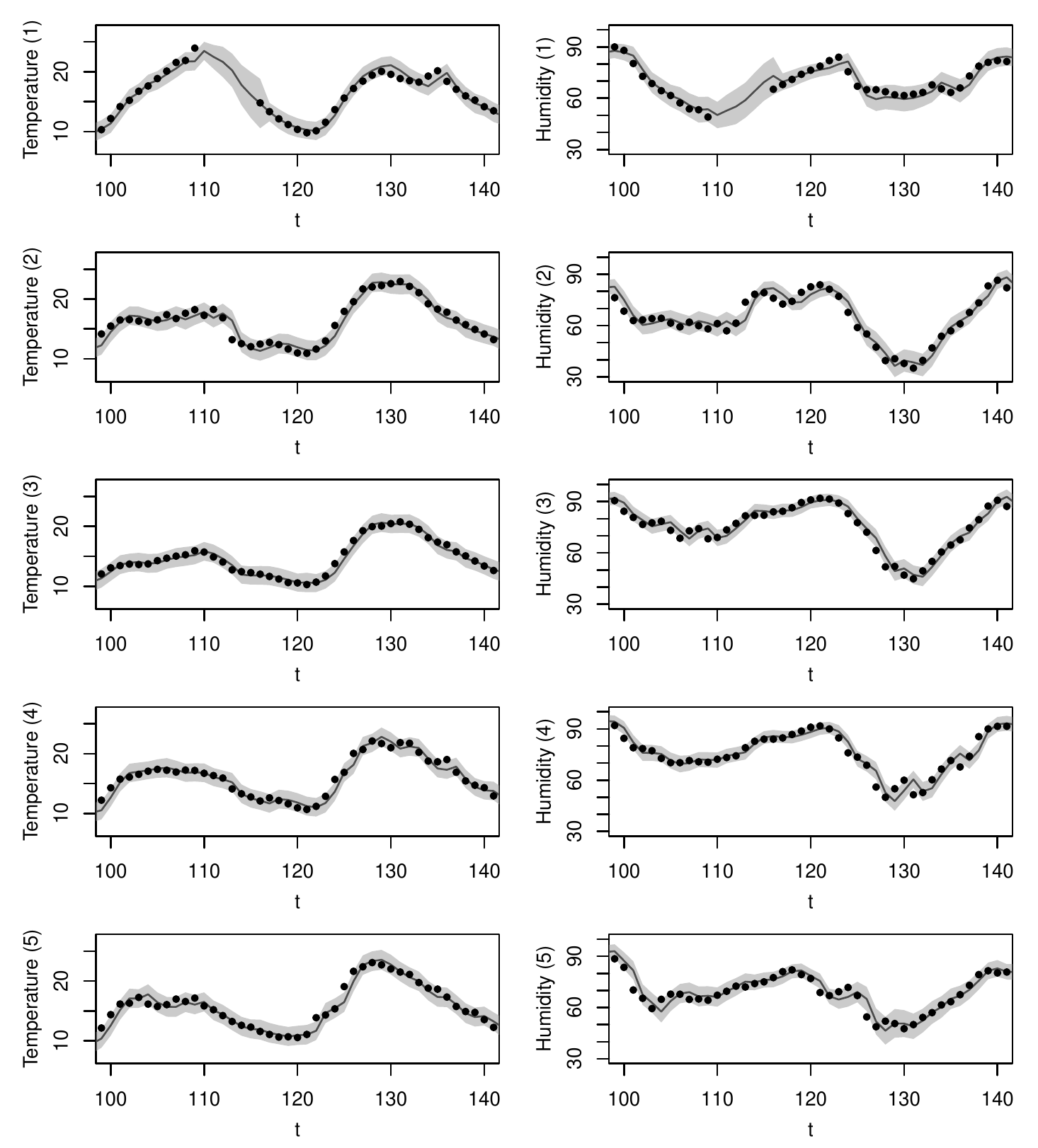}
\caption{One-step ahead forecast means (------) and 95\% credible
  intervals, at each location (1--5) over time. The observations are
  indicated ($\bullet$). The observation period is from 12th July 2017
  08:00:00 to 14th July 2017 00:00:00.}
\label{forecast}
\end{figure}

\begin{figure}[p]
\centering
\includegraphics[width=1\textwidth,height=0.8\textheight]{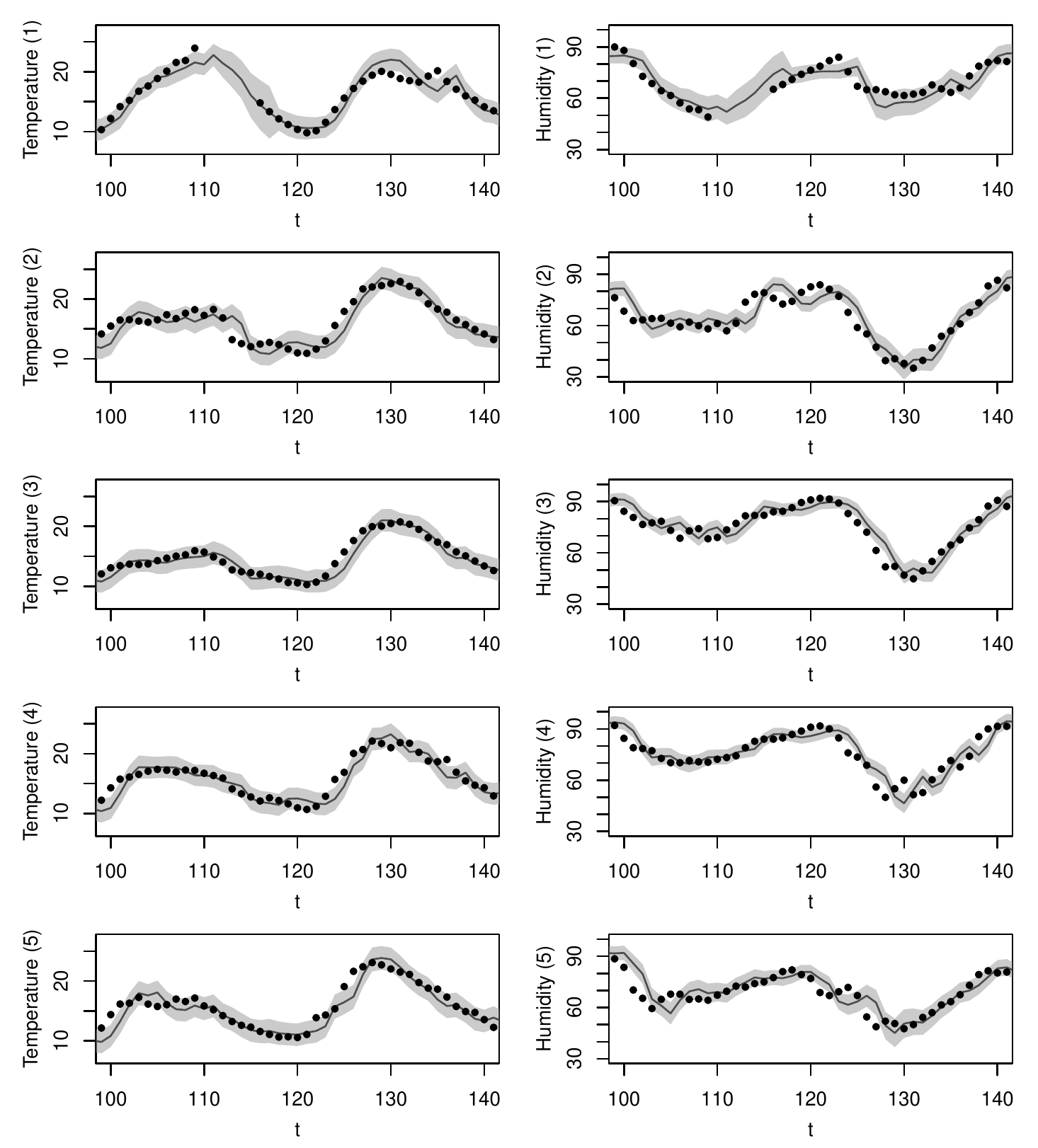}
\caption{Two-step ahead forecast means (------) and 95\% credible
  intervals, at each location (1--5) over time. The observations are
  indicated ($\bullet$). The observation period is from 12th July 2017
  08:00:00 to 14th July 2017 00:00:00.}
\label{forecast2}
\end{figure}

We assess the validity of the proposed model by comparing observed
data with their model-based within-sample posterior predictive
distributions and with model-based out-of-sample forecast
distributions. Simulation methods can be used to construct these
distributions and details on how to generate draws from them is
provided in Appendix~\ref{app:prediction}. Figure~\ref{prediction}
shows discrepancies between observations and their within-sample
predictive distribution over the first 500 hours at each of the 5
locations. These distributions are characterised by their mean and
95\% credible interval. It is clear that the mean difference at each
time-location combination is small and that a mean difference of zero
is plausible (the 95\% credible intervals include zero).  Similar
results were obtained for the full data set (not shown).
Figure~\ref{forecast} shows the mean and 95\% credible interval at
each location for the one-step ahead forecast. The times displayed
were chosen at random over a two day period and, for comparison
purposes, the observations at these times are also shown.
Unsurprisingly forecast uncertainty grows during periods of prolonged
missingness. The figure shows that observations typically lie within
the forecast interval and that the model-based one-step forecast
distribution is consistent with the observed data.
Figure~\ref{forecast2} shows the mean and 95\% credible interval at
each location for the two-step ahead forecast. Similar to the one-step
forecasts, this figure shows that these forecast distributions are
consistent with the data but, of course, have larger uncertainty.

\begin{figure}[p]
\centering
\includegraphics[width=1\textwidth,height=0.8\textheight]{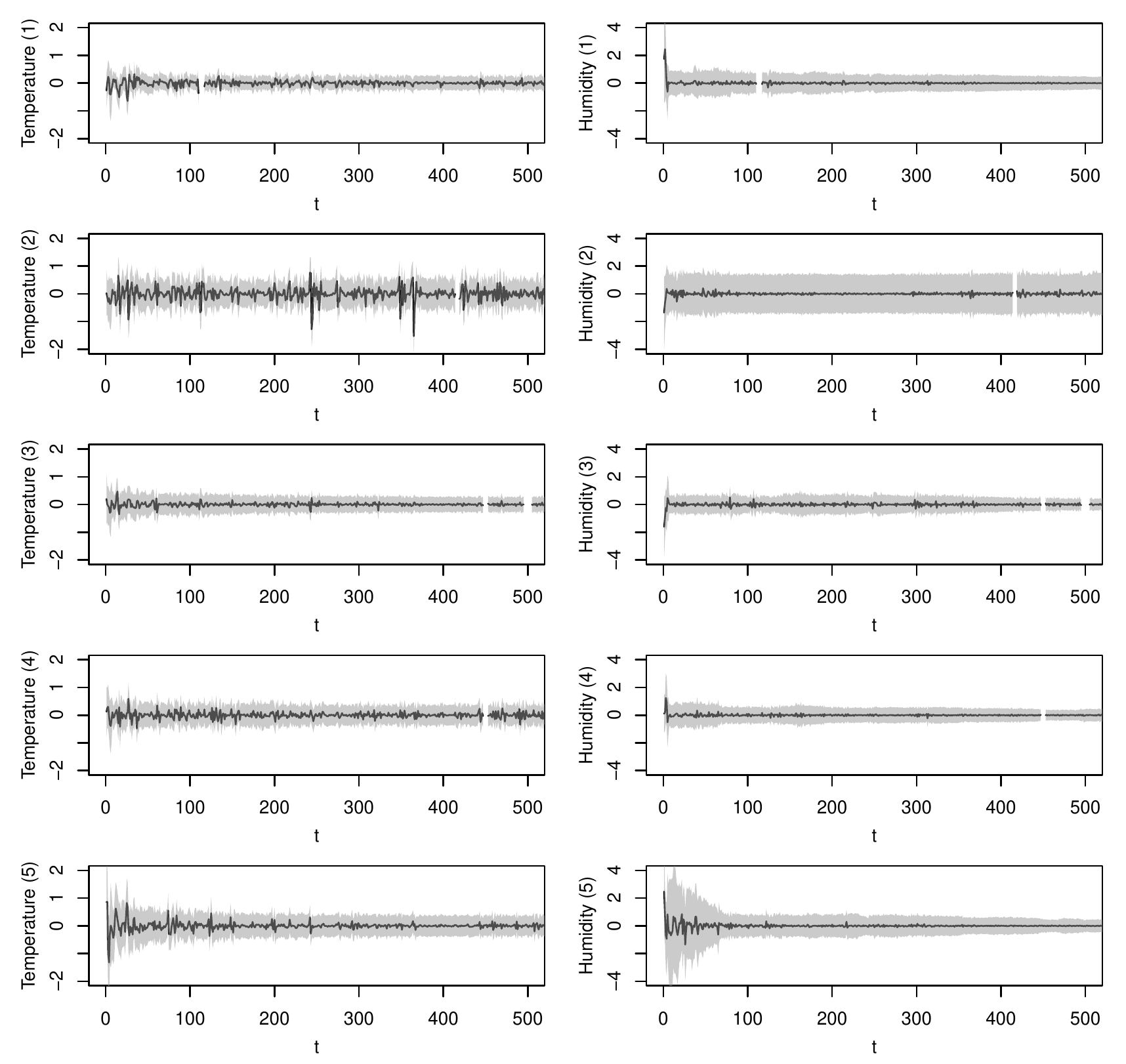}
\caption{Mean (------) and 95\% credible intervals for the difference
  between the within-sample predictive and the observations, at each
  location (1--5) over time. The observation period is from 8th July
  2017 04:00:00 to 29th July 2017 00:00:00.}
\label{prediction}
\end{figure}

\section{Discussion}
\label{disc}
We have developed and fitted a spatio-temporal model to around six
months of data on hourly temperature and humidity values at five
locations in the North East of England. The data were obtained from a
sensor network providing streaming data on environmental variables
such as climate, pollution and traffic flow, held at the Newcastle
Urban Observatory. The model we use for observed seasonality in
temperature is a dynamic linear model (DLM) whose observation equation
takes the form of a sinusoid, with time varying amplitude and phase
described by the system equation. We capture the observed linear
relationship between humidity and temperature via a conditional DLM in
which humidity is regressed on temperature.  Also spatial consistency
at nearby sites is accounted for by adding a Gaussian process in the
system equations.

Our primary goal is real time forecasting of temperature and humidity.
To this end, we have developed a sequential Monte Carlo (SMC)
algorithm which updates the parameter posterior as each measurement
becomes available. The tractability of the observed data likelihood
allows us to construct the SMC algorithm using an iterated batch
importance sampling (IBIS) scheme, first introduced by
\cite{Chopin02}. The IBIS scheme tries to deal with particle
degeneracy by employing a resample-move step which allows the particle
set to be rejuvenated by moving each particle through a
Metropolis-Hastings kernel that leaves the target posterior invariant.
The computational cost of this step increases as the algorithm runs,
due to the time taken to calculate the observed data likelihood at
each particle, as more data is included. This problem is made much
more accute by the long length of the observed time series and the
high dimension of the parameter space and this makes the algorithm
unusable as an on-line algorithm.  To circumvent this issue, we have
modified the resample-move step in two ways.  First, we use a sequence
of observation windows and calculate the observed data likelihood for
the data within the window. As the data in each window are included,
the parameter posterior (at the start of the window) is approximated
using a kernel density estimate and then updated using the observed
data likelihood for the window. This places an upper bound on the
computational cost. We looked the effect of the choice of window
length on computational efficiency and posterior accuracy and found
that reasonable posterior accuracy can be achieved for modest window
length.  Finally, we speed up the algorithm by using a fully parallel
implementation which divides the particles into batches and performs
the resampling step locally, for each batch. We term the resulting
scheme \emph{online IBIS} and find that for our data set, an
observation (consisting of both temperature and humidity hourly
averages at each of five locations) can be assimilated in around 3
minutes on average, with this average time dominated by the
rejuvenation steps. One-step and two-step forecast distributions can
then be determined very quickly. Given that observations arrive every
hour, this makes the scheme entirely feasible for use in real time.

  This work can be extended in a number of ways. For example,
  covariate information such as altitude, distance from the coast and
  wind direction/speed could be included in the model. Unfortunately
  this information is not currently available. Developing a joint
  model for all sensor streams, which would also include pollution
  data and traffic data, is also of interest. However, fitting models
  of multiple heterogeneous sensors is likely to require further
  methodological development of the inference scheme considered here.

\begingroup
\let\itshape\upshape

\bibliographystyle{unsrt}
\bibliography{references}   

\endgroup

\appendix

\section{Appendix}

\subsection{Forward filter}
\label{sec:ff}
To simplify notation we consider the spatial temperature model and
drop $x$. Given the form of the observation model in (\ref{obsmod}),
we have that
\begin{equation}
\begin{split}
\vec{X}_{t_i}^o &= \tilde{\vec{F}}_{t_i}\vec{\theta}_{t_i}+\tilde{\vec{v}}_{i},\qquad \tilde{\vec{v}}_{i}\indep N(\vec{0},\tilde{\vec{V}}), \\  
\vec{\theta}_{t_i}&=\vec{\theta}_{t_{i-1}}+\tilde{\vec{w}}_{i},\qquad \tilde{\vec{w}}_{i}\indep N(\vec{0},\tilde{\vec{W}}),
\end{split}
\label{DLMobs} 
\end{equation}
where $\tilde{\vec{F}}_{t_i}=\vec{P}_{t_i}\vec{F}_{t_i}$, $\tilde{\vec{V}}=\vec{P}_{t_i}\textrm{diag}(V^{1},\ldots,V^{\L})\vec{P}_{t_i}^T$ and $\tilde{\vec{W}}=k_i^2\textrm{diag}(\vec{W}^{1},\ldots,\vec{W}^{\L})+\vec{K}$. 
Since the parameters $\vec{\phi}$ remain fixed throughout this section, we drop them from
the notation where possible. Now suppose that $\vec{\theta}_{t_1}\sim N(\vec{m},\vec{C})$ \emph{a priori} and recall that $t_1=0$. The observed data likelihood increments 
$\pi(\vec{x}^o_{t_i}|\vec{x}^o_{0:t_{i-1}})$, and hence the full
observed data likelihood $\pi(\vec{x}^o_{0:t_n}|\vec{\phi})$, can be
obtained from the forward filter described in Algorithm~\ref{ff}.

\begin{algorithm}[t]
\caption{Forward filter}\label{ff}
\begin{enumerate}
\item Initialisation ($i=1$). Compute
  $\pi(\vec{x}^o_{t_1})=N(\vec{x}_{t_{1}}\,;\, \tilde{\vec{F}}_{t_1}\vec{m}\,,\,\tilde{\vec{F}}_{t_1}\vec{C}\tilde{\vec{F}}_{t_1}^T+\tilde{\vec{V}})$. The
  posterior at time $t_1=0$ is therefore $\vec{\theta}_{t_1}|\vec{x}_{t_1}^o\sim
  N(\vec{m}_{t_1},\vec{C}_{t_1})$, where
\begin{align*}
\vec{m}_{t_1} &= \vec{m}+\vec{C}\tilde{\vec{F}}_{t_1}^T(\tilde{\vec{F}}_{t_1}\vec{C}\tilde{\vec{F}}_{t_1}^T+\tilde{\vec{V}})^{-1}(\vec{x}_{t_1}^o-\tilde{\vec{F}}_{t_1}\vec{m}) \\
\vec{C}_{t_1} &= \vec{C}-\vec{C}\tilde{\vec{F}}_{t_1}^T(\tilde{\vec{F}}_{t_1}\vec{C}\tilde{\vec{F}}_{t_1}^T+\tilde{\vec{V}})^{-1}\tilde{\vec{F}}_{t_1}\vec{C}\,.
\end{align*}
Store the values of $\vec{m}_{t_1}$, $\vec{C}_{t_1}$ and $\pi(\vec{x}^o_{t_1})$.

\item For $i=2,\ldots,n$,
\begin{itemize}
\item[(a)] Prior at $t_{i}$. Using the system equation, we have that $\vec{\theta}_{t_i}|\vec{x}^o_{0:t_{i-1}}\sim N(\vec{m}_{t_{i-1}},\vec{C}_{t_{i-1}}+\tilde{\vec{W}})$.
\item[(b)] One step forecast. Using the observation equation, we have that 
\[
\vec{X}_{t_i}^o|\vec{x}_{0:t_{i-1}}^o\sim N\{\tilde{\vec{F}}_{t_i}\vec{m}_{t_{i-1}},\tilde{\vec{F}}_{t_i}(\vec{C}_{t_{i-1}}+\tilde{\vec{W}})\tilde{\vec{F}}_{t_i}^T+\tilde{\vec{V}}\}.
\]
Compute the observed data likelihood increment
\begin{align*}
\pi(\vec{x}^o_{t_i}|\vec{x}^o_{0:t_{i-1}})&=N\{\vec{x}^o_{t_i}\,;\,\tilde{\vec{F}}_{t_i}\vec{m}_{t_{i-1}}\,,\,\tilde{\vec{F}}_{t_i}(\vec{C}_{t_{i-1}}+\tilde{\vec{W}})\tilde{\vec{F}}_{t_i}^T+\tilde{\vec{V}}\}.
\end{align*}
\item[(c)] Posterior at $t_i$. Combining the distributions in (a) and (b) gives the joint 
distribution of $\vec{\theta}_{t_i}$ and $\vec{X}^o_{t_i}$ (conditional on $\vec{x}_{0:t_{i-1}}$) as
\[
\begin{pmatrix}
	\vec{\theta}_{t_i} \\	
	\vec{X}^o_{t_i}
	\end{pmatrix}\sim N\left\{\begin{pmatrix}
	\vec{m}_{t_{i-1}} \\
	\tilde{\vec{F}}_{t_i}\vec{m}_{t_{i-1}} 	
	\end{pmatrix}\,,\, \begin{pmatrix}
	 \vec{C}_{t_{i-1}}+\tilde{\vec{W}} & (\vec{C}_{t_{i-1}}+\tilde{\vec{W}})\tilde{\vec{F}}_{t_i}^T  \\
	\tilde{\vec{F}}_{t_i}(\vec{C}_{t_{i-1}}+\tilde{\vec{W}}) & \tilde{\vec{F}}_{t_i}(\vec{C}_{t_{i-1}}+\tilde{\vec{W}})\tilde{\vec{F}}_{t_i}^T+\tilde{\vec{V}}  	 
	\end{pmatrix} \right \} 
\]
and therefore $\vec{\theta}_{t_i}|\vec{x}^o_{0:t_i}\sim N(\vec{m}_{t_i},\vec{C}_{t_i})$, where
\begin{align*}
\vec{m}_{t_i} &= \vec{m}_{t_{i-1}}+ (\vec{C}_{t_{i-1}}+\tilde{\vec{W}})\tilde{\vec{F}}_{t_i}^T\{\tilde{\vec{F}}_{t_i}(\vec{C}_{t_{i-1}}+\tilde{\vec{W}})\tilde{\vec{F}}_{t_i}^T+\tilde{\vec{V}}\}^{-1}(\vec{x}^o_{t_i}-\tilde{\vec{F}}_{t_i}\vec{m}_{t_{i-1}}) \\
\vec{C}_{t_i} &= \vec{C}_{t_{i-1}}+\tilde{\vec{W}} - (\vec{C}_{t_{i-1}}+\tilde{\vec{W}})\tilde{\vec{F}}_{t_i}^T\{\tilde{\vec{F}}_{t_i}(\vec{C}_{t_{i-1}}+\tilde{\vec{W}})\tilde{\vec{F}}_{t_i}^T+\tilde{\vec{V}}\}^{-1}\tilde{\vec{F}}_{t_i}(\vec{C}_{t_{i-1}}+\tilde{\vec{W}})  \,.
\end{align*}
Store the values of $\vec{m}_{t_i}$, $\vec{C}_{t_i}$ and $\pi(\vec{x}^o_{t_i}|\vec{x}^o_{0:t_{i-1}})$.
\end{itemize}
\end{enumerate}
\end{algorithm}

\subsection{Within-sample predictions and out-of-sample forecasts}
\label{app:prediction}
In order to compute within-sample predictions, the smoothing density
$\pi(\vec{\theta}_{0:t_n}|\vec{x}^o_{0:t_n},\vec{\phi}_x)$ is
required. Draws from this density can be readily obtained by using a
backward sampler that recursively draws from
\begin{equation}\label{syspred}
\pi(\vec{\theta}_{t_i}|\vec{\theta}_{t_{i+1}},\vec{x}^o_{0:t_i},\vec{\phi}_x)= N\{\vec{\theta}_{t_i};\,\vec{m}_{t_i}+\vec{B}_{t_i}(\vec{\theta}_{t_{i+1}}-\vec{m}_{t_i})\,,\, 
\vec{C}_{t_i}-\vec{B}_{t_i}\vec{R}_{t_{i+1}}\vec{B}_{t_i}^{T}\},
\end{equation}
where $\vec{B}_{t_i}=\vec{C}_{t_i}\vec{R}_{t_{i+1}}^{-1}$ and
$\vec{R}_{t_{i+1}}=\vec{C}_{t_i}+\tilde{\vec{W}}$; see, for example, \cite{West99}. Hence, given an
equally weighted sample $\{\vec{\phi}_x^{1:N}\}$ from the marginal
posterior $\pi(\vec{\phi}_x|\vec{x}^o_{0:t_n})$, we can integrate over
parameter uncertainty to generate draws from the within-sample system
posterior predictive density
$\pi(\vec{\theta}_{0:t_n}|\vec{x}^o_{0:t_n})$ by recursively drawing
from (\ref{syspred}) for each particle $\vec{\phi}_x^{(k)}$ (and the
associated quantities $\vec{m}_{t_i}^{(k)}$, $\vec{C}_{t_i}^{(k)}$
generated by the forward filter). Subsequently, the within-sample
observation posterior predictive density
$\pi(\vec{x}_{0:t_n}|\vec{x}^o_{0:t_n})$ can be sampled by drawing
\[
\vec{X}_{t_i}^{(k)}|\vec{\theta}_{t_i}^{(k)},\vec{\phi}_x^{(k)}
\sim N(\vec{F}_{t_i}\vec{\theta}_{t_i}^{(k)}\,,\,\vec{V}^{(k)}),
\qquad i=1,\ldots,n,\quad k=1,\ldots,N.
\]

Out-of-sample system and observation forecast distributions can be obtained by again exploiting the linear Gaussian structure of the DLM. Given an equally weighted sample $\{\vec{\phi}_x^{1:N}\}$ from the marginal posterior $\pi(\vec{\phi}_x|\vec{x}^o_{0:t_n})$, samples from $\pi(\vec{\theta}_{t_{n+1}}| \vec{x}^o_{0:t_n})$ and $\pi(\vec{x}_{t_{n+1}}|\vec{x}^o_{0:t_n})$ can be obtained by recursively drawing
\begin{align*}
\vec{\theta}_{t_{n+1}}^{(k)}| \vec{\phi}_x^{(k)} 
&\sim N(\vec{m}_{t_n}^{(k)}\,,\, \vec{C}_{t_n}^{(k)}+\tilde{\vec{W}}^{(k)}),\qquad k=1,\ldots,n\\
\vec{x}_{t_{n+1}}^{(k)}|\vec{\phi}_x^{(k)} 
&\sim N\{\vec{F}_{t_{n+1}}\vec{m}_{t_n}^{(k)}\,,\,\vec{F}_{t_{n+1}}(\vec{C}_{t_n}^{(k)}+\tilde{\vec{W}}^{(k)}) \vec{F}_{t_{n+1}}^T+\vec{V}^{(k)}\},\qquad k=1,\ldots,n.
\end{align*}

\subsection{Model selection}
\label{app:modelselec}
As noted in Section~\ref{sec:tdlm}, seasonality in the marginal DLM
can be accounted for in two ways. A sinusoid can be specified in the
observation equation, with a system equation describing the evolution
of the parameters governing the amplitude and phase. Alternatively, a
Fourier form structure can be used in the system equation where the
appropriate number of harmonics must be specified by the practitioner.
Our joint model consists of a marginal DLM for temperature and a
conditional DLM for humidity given tempertaure. This induces a
marginal DLM for humidity with the same form as that for temperature.
We therefore consider three candidate spatial DLMs for modelling
temperature and humidity data marginally: 1. sinusoidal form DLM
(sDLM); 2. Fourier form DLM with 1 harmonic (FDLM1); 3. Fourier form
DLM with 2 harmonics (FDLM2). Choosing between these competing models
is possible via computation of the Bayes factor
\citep{kass95,schnatter95}, which, under the assumption of equal prior
probability for two competing models, say $M1$ and $M2$, is defined as
the ratio of the evidence given $M1$, and that given $M2$. The Bayes
factor based on temperature data is therefore
\[
BF=\frac{p(\bm{x}_{0:t_{n}}^o | M1)}{p(\bm{x}_{0:t_{n}}^o | M2)}
\] 
with a similar form for the humidity data Bayes factor. Note that
$BF<1$ suggests the data support $M2$. Equation (\ref{ev}) gives an
estimate of the evidence as a by-product of the IBIS scheme.

\begin{figure}[H]
\centering
\includegraphics[width=1\textwidth,height=0.36\textheight]{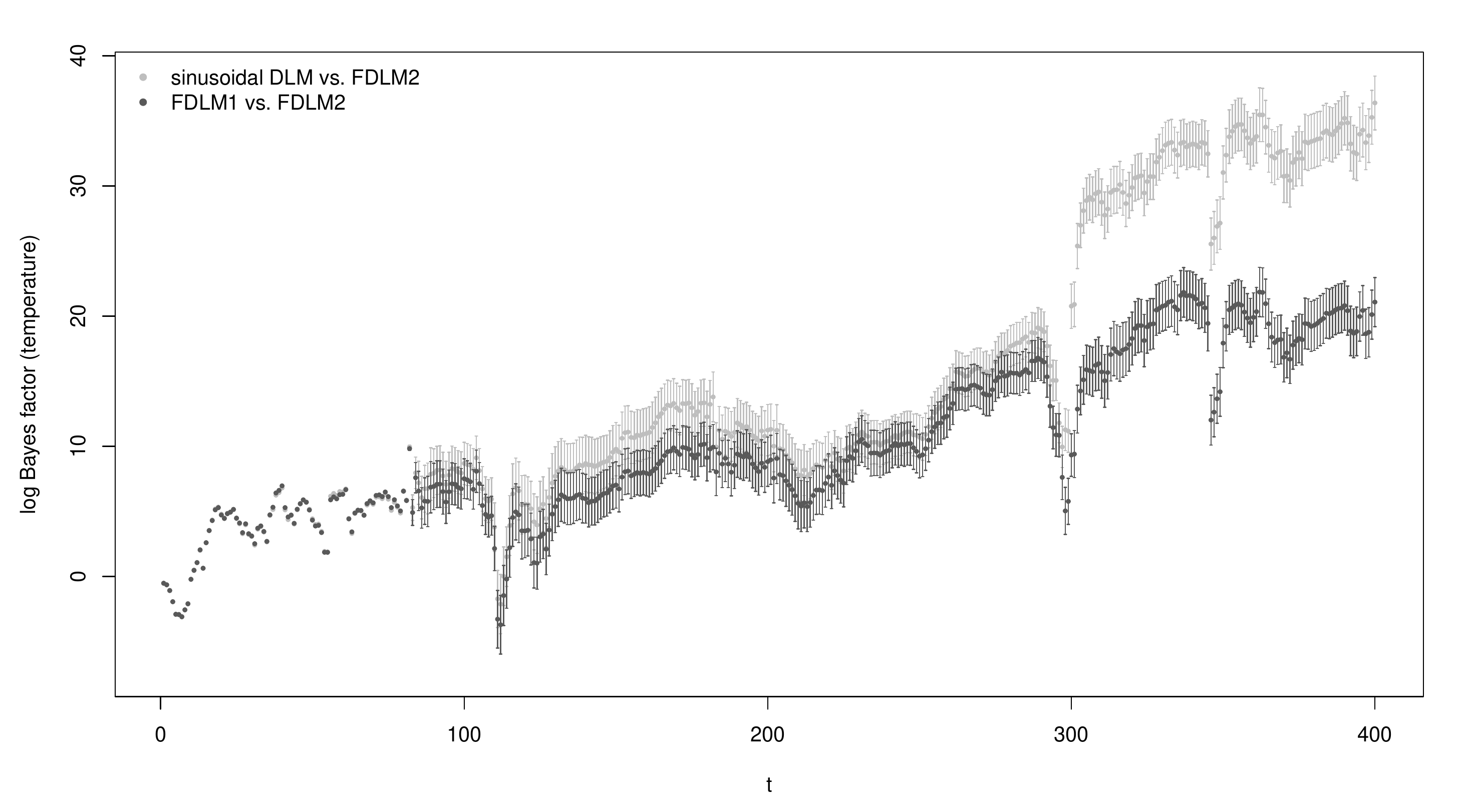}

\includegraphics[width=1\textwidth,height=0.36\textheight]{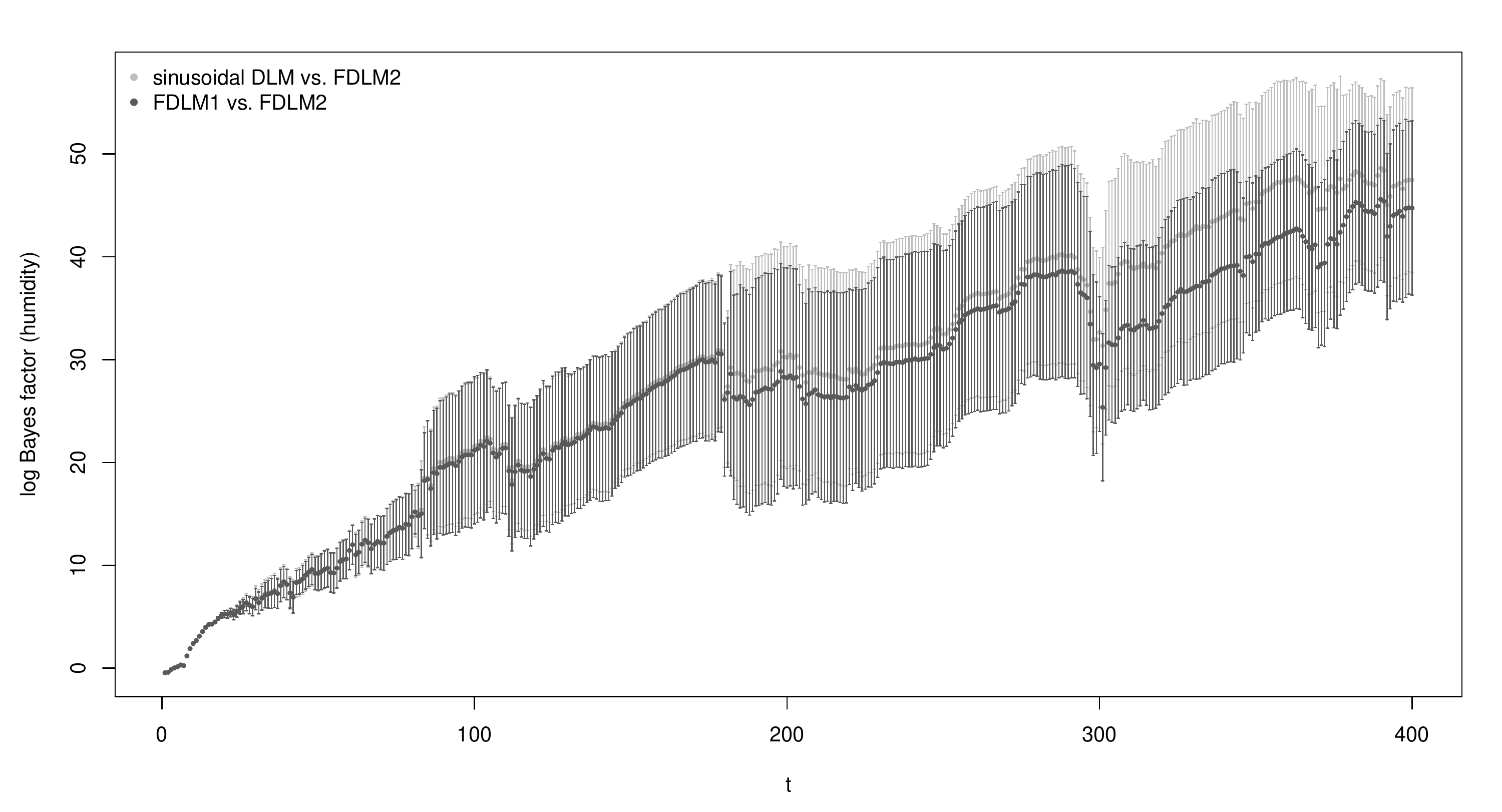}
\caption{Mean and 95\% credible interval of the log Bayes factor comparing sDLM against FDLM2 and FDLM1 against FDLM2, over time. (Top: temperature models; bottom: humidity models.)}
\label{modelselc}
\end{figure}

Unfortunately, the size of the observed dataset precludes calculation
of the Bayes factor using all measurements at all sites. Therefore, to
guide our modelling approach we chose three of the five locations at
random and then 400 consecutive observations (starting at a random
observed time) at these locations. The evidence for each model was
determine using the full IBIS scheme on these data with a serial
multinomial resampling step for each model, using $N=10^7$ particles.
To account for Monte Carlo error, we repeat this process 30 times.
Taking FDLM2 as a baseline for comparison, we compute Bayes factors
for sDLM vs FDLM2 and FDLM1 vs FDLM2. Figure~\ref{modelselc} shows the
mean $\log BF$ value (and 95\% credible interval) based on data
$\bm{x}_{0:t}^o$ and $\bm{y}_{0:t}^o$ against $t$. For the marginal
temperature DLM it is clear that FDLM2 is the least favoured model.
Furthermore, for $t>80$, the log Bayes factors corresponding to the
sinusoidal form DLM against FDLM2 are always strictly greater than
those corresponding to FDLM1 against FDLM2. For the marginal humidity
DLM, there is little difference in overall fit between the sinusoidal
form DLM and FDLM1. Given that computational cost scales as
$1:1.1:1.3$ for $DLM:FDLM1:FDLM2$, we conclude that the sinusoidal
form DLM offers the best compromise between model fit and
computational efficiency.

\end{document}